\documentclass[12pt]{article}
\pdfoutput=1

\usepackage{amssymb,amsmath,mathrsfs,epstopdf,slashed,color}
\usepackage{tikz}
\usetikzlibrary{matrix}
\usetikzlibrary{positioning}

\usepackage{graphicx}
\usepackage{placeins}
\usepackage{hyperref}
\usepackage{cite}

\allowdisplaybreaks[1]

\makeatletter

\@addtoreset{equation}{section}
\makeatother

\newcommand{\vol}{\mathrm{vol}}

\title{\bf Linking Dark Matter to Dark Energy in A String Theory Scenario}
\author{S.-H. Henry Tye, and Sam S.C. Wong}

\begin{document}

\begin{titlepage}

\setcounter{page}{0}
  
\begin{flushright}
 \small
 \normalsize
\end{flushright}

\vskip 3cm
\begin{center}

{\Large \bf Linking Light Scalar Modes with A Small Positive Cosmological Constant in String Theory}  

\vskip 2cm
  
{\large S.-H. Henry Tye${}^{1,2}$, and Sam S.C. Wong${}^1$}
 
 \vskip 0.6cm

 ${}^1$ Jockey Club Institute for Advanced Study and Department of Physics, Hong Kong University of Science and Technology, Hong Kong\\
 ${}^2$ Laboratory for Elementary-Particle Physics, Cornell University, Ithaca, NY 14853, USA

 \vskip 0.4cm

Email: \href{mailto: iastye@ust.hk, scswong@ust.hk}{iastye at ust.hk, scswong at ust.hk}

\vskip 1.0cm
  
\abstract{\normalsize
Based on the studies in Type IIB string theory phenomenology, we conjecture that a good fraction of the meta-stable de Sitter vacua in the cosmic stringy landscape tend to have a very small cosmological constant $\Lambda$ when compared to either the string scale $M_S$ or the Planck scale $M_P$, i.e., $\Lambda \ll M_S^4 \ll M_P^4$. These low lying de Sitter vacua tend to be accompanied by very light scalar bosons/axions. Here we illustrate this phenomenon with the bosonic mass spectra in a set of Type IIB string theory flux compactification models. We conjecture that small $\Lambda$ with light bosons is generic among de Sitter solutions in string theory; that is, the smallness of $\Lambda$ and the existence of very light bosons (may be even the Higgs boson) are results of the statistical preference for such vacua in the landscape. We also discuss a scalar field $\phi^3/\phi^4$ model to illustrate how this statistical preference for a small $\Lambda$ remains when quantum loop corrections are included, thus 
bypassing the radiative instability problem.

}
  
\vspace{1cm}
\begin{flushleft}
 \today
\end{flushleft}
 
\end{center}
\end{titlepage}

\setcounter{page}{1}
\setcounter{footnote}{0}

\tableofcontents

\parskip=5pt

\section{Introduction}

Cosmological data strongly indicates that our universe has a vanishingly small positive cosmological constant $\Lambda$ (or vacuum energy density) as the dark energy,
\begin{equation}
\Lambda \sim 10^{-122}M_P^4
 \label{L1}
\end{equation}
where the Planck mass  $M_P = G_N^{-1/2}\simeq 10^{19}$ GeV.  The smallness of $\Lambda$ is a major puzzle in physics.
In general relativity, $\Lambda$ is a free arbitrary parameter one can introduce, so its smallness can be accommodated but not explained within quantum field theory. 
On the other hand, string theory has only a single parameter, namely the string scale $M_S=1/\sqrt{2 \pi \alpha' }$, so everything else should be calculable for each string theory solution. String theory has 9 spatial dimensions, 6 of them must be dynamically compactified to describe our universe. Since both $M_P$ and $\Lambda$ are calculable, $\Lambda$ can be determined in terms of $M_P$ dynamically in each local minimum compactification solution. This offers the possibility that we may find an explanation for a very small positive $\Lambda$. This happens if a good fraction of the meta-stable deSitter (dS) vacua in the landscape tend to have a very small $\Lambda$, as is the case in the few studies in flux compactification in string theory \cite{Sumitomo:2012vx,Danielsson:2012by,Sumitomo:2013vla}.

There are many studies performed in the search of meta-stable dS vacua in string theory \cite{Douglas:2006es}.  Such searches must be elaborate enough to \\
(1) stabilize all the moduli via flux compactification, in which the fluxes are quantized \cite{Bousso:2000xa,Giddings:2001yu}. For multiple moduli cases, statistical analysis suggests that the probability that all moduli are stabilized (i.e., with semi-positive mass-squared) is Gaussianly suppressed \cite{Aazami:2005jf,Dean:2006wk,Borot:2010tr,Marsh:2011aa,Chen:2011ac}. If we uplift an anti-deSitter (AdS) vacuum to a dS vacuum, one can imagine that stability is harder to maintain as the vacuum energy grows, suggesting that there are fewer of them compared to AdS solutions;\\
(2) bypass the no-go theorems that forbid dS vacua in model-buildings with positive Euler number $\chi$ or without orientifold planes \cite{Maldacena:2000mw,Hertzberg:2007wc,Wrase:2010ew,Caviezel:2009tu,Danielsson:2011au,Shiu:2011zt,Danielsson:2012et,Junghans:2016uvg}.

To simplify the discussion, let us focus on flux compactification of Type IIB theory to 4 dimensional spacetime. 
Fortunately, there are examples where the existence of dS vacua is likely, e.g., the KKLT scenario \cite{Kachru:2003aw}, the large volume scenario \cite{Balasubramanian:2005zx}, the K\"ahler uplift scenario \cite{Westphal:2006tn,Rummel:2011cd} and the non-geometric flux scenario \cite{deCarlos:2009fq,deCarlos:2009qm,Blaback:2013ht,Hassler:2014mla,Danielsson:2015tsa}. 
Start with the four-dimensional low energy (supergravity) effective potential $V (F_i, \phi_j)$, where  $F_i$ are the 4-form field strengths and $\phi_j$ are the complex moduli (and dilaton) describing the size and shape of the compactified manifold as well as the coupling.
It is known that the field strengths $F_i$ in flux compactification in string theory take only quantized values at the local minima \cite{Bousso:2000xa}. 
In the search of classical minima, this flux quantization property allows us to rewrite $V( F_i, \phi_j)$ as a function of  the quantized values $n_i$ of the fluxes present, 
$$V( F_i, \phi_j) \rightarrow V(n_i, \phi_j), \quad \quad i=1,2,...,N, \quad j=1,2,..., K.$$
Since string theory has no continuous free parameter, there is no arbitrary free parameter in $V(n_i, \phi_j)$, though it does contain (in principle) calculable quantities like $\alpha'$ corrections, loop and non-perturbative corrections, and geometric quantities like Euler index $\chi$ etc.. 

For a given set of discrete flux parameters $\{n_i\}$, we can solve $V(n_i, \phi_j)$ for its meta-stable (classically stable) vacuum solutions via finding the values $\phi_{j, {\rm min}} (n_i)$ at each solution and determine its vacuum energy density $\Lambda=\Lambda(n_i, \phi_{j, {\rm min}} (n_i))=\Lambda (n_i)$.  Since we are considering the physical $\phi_j$, it is the physical $\Lambda$ we are determining.  Since a typical flux parameter $n_i$ can take a large range of integer values, we may simply treat each $n_i$ as an independent random variable with some  distribution $P_i(n_i)$. Collecting all such solutions, we can next find the probability distribution $P(\Lambda)$ of $\Lambda$ of these meta-stable solutions as we sweep through all the flux numbers $n_i$.  That is putting $P_i(n_i)$ and $\Lambda (n_i)$ together yields $P(\Lambda)$, 
$$P(\Lambda)= \sum_{n_i} \delta(\Lambda - \Lambda (n_i))  \Pi_i P_i(n_i), $$ so $\sum_{n_i} P_i(n_i) =1$ for each $i$ implies that $\int P(\Lambda) d \Lambda =1$.  For large enough ranges for $n_i$, we may treat each $P_i(n_i)$ as a continuous function over an appropriate range of values. 
This strategy of doing statistics is different from that of Ref\cite{Denef:2004ze,Denef:2004cf} where the superpotential $W$ and its derivatives $DW$ and $DDW$ are treated as independent random variables but the meta-stable minima are not solved in terms of flux parameters. 

Simple probability properties show that $P(\Lambda)$ easily peaks and diverges at $\Lambda=0$ \cite{Sumitomo:2012wa}, implying that a small $\Lambda$ is statistically preferred. For an exponentially small $\Lambda$, the statistical preference for $\Lambda \simeq 0$ has to be overwhelmingly strong, that is, $P(\Lambda)$ has to diverge (i.e., peak) sharply at $\Lambda =0$. Such an analysis has been applied to the K\"ahler uplift scenario \cite{Rummel:2011cd}, where $P(\Lambda)$ is so peaked at $\Lambda=0$ that the the median $\Lambda$ matches the observed $\Lambda$ (\ref{L1}) if the number of complex structure moduli $h^{2,1} \sim{ \cal O} (100)$ \cite{Sumitomo:2012vx}.  Such a value for $h^{2,1}$ is quite reasonable for a typical manifold considered in string theory. That is, an overwhelmingly large number of meta-stable vacua have an exponentially small $\Lambda$, so statistically, we should end up in one of them. In other words, a very small $\Lambda$ is quite natural. The preference for a very small $\Lambda$ has also been observed in the racetrack scenario \cite{Sumitomo:2013vla}.  In the non-geometric flux scenario \cite{Danielsson:2012by}, it is also found that dS vacua are surprisingly rare, and they appear mostly with small values of $\Lambda$. 
This leads us to conjecture that
 
{\bf Substantial regions of the cosmic stringy deSitter landscape is dominated by meta-stable vacua with $\Lambda \ll M_S^4$.}
 
If true, this may (i) provide an explanation why the observed $\Lambda$ is so small, and (2) after inflation, why the universe is not trapped in a relatively high $\Lambda$ vacuum. 
A few comments are in order here: 

$\bullet$ The existence of the landscape is crucial for this explanation why a very small $\Lambda$ is natural.
 It remains to be seen how large these regions are in the whole cosmic landscape.

$\bullet$ 
Recall the Bousso-Polchinski scenario \cite{Bousso:2000xa}. If there are a dozen or more independent flux parameters present, the allowed $\Lambda$ values can form a sufficiently dense ``discretuum" so the spacings between neighboring values are comparable to the observed $\Lambda$. That is, the observed $\Lambda$ can be easily accommodated. \footnote{See \cite{Bachlechner:2015gwa} for a similar idea in an axion landscape.}
However, not all choices of flux values yield meta-stable vacua, while multiple solutions may appear for a single choice of fluxes. 
As a result, we see that only a tiny set of fluxes yield dS solutions. In the models studied, we find that most of the dS solutions have $\Lambda \simeq 0$. Looking at the $\phi^3/\phi^4$ model in Sec. 2 and Ref\cite{Sumitomo:2012wa}, we see that this statistical preference for $\Lambda \simeq 0$ is a simple consequence of elementary probability theory. In fact, this preference $\Lambda \simeq 0$ is absent when couplings are absent. This suggests the generic property that the peaking of $P(\Lambda)$ at $\Lambda=0$ is enhanced (or at least not suppressed) by more couplings among the moduli/fields. This tendency offers the hope that the simple cases studied so far do reflect the actual situation, when couplings among fields are highly non-trivial, but unfortunately more difficult to analyze.

$\bullet$ We observe that the peaking of $P(\Lambda)$ (i.e., its divergent behavior) at $\Lambda=0$ is relatively insensitive to the particular forms of the input probability distributions $P_i(n_i)$ \cite{Sumitomo:2012wa}; it is the functional form of  $V(n_i, \phi_j)$, hence $\Lambda (n_i)$, that is important. For a dense enough discretuum of $\Lambda$, the discrete flux parameters $n_i$ may be treated as random variables with continuous (relatively smooth) values over some appropriate ranges. 

$\bullet$ In usual quantum field theory, even if we include up to the $n$th radiative loop effect to obtain a very small $\Lambda_n$ that is comparable to the observed value (\ref{L1}), the $(n+1)$th loop correction tends to shift $\Lambda_n$ by an amount much bigger than it, i.e., $|\Lambda_{n+1}| \simeq |\delta \Lambda| \gg \Lambda_n$. To have a very small $\Lambda_{n+1}$, we have to fine-tune the input couplings/parameters. This property is known as radiative instability, which is a main stumbling block in understanding why, in the absence of fine-tuning, the physical $\Lambda$ can be so small. Since string theory has no free couplings/parameters to be fine-tuned, one may naively think this radiative instability problem may be more severe in string theory. However, the cosmic stringy landscape offers a way out. Here we sweep through all allowed values of the couplings/parameters in the low energy effective potential and find the probability distribution $P(\Lambda)$ of $\Lambda$. As long as $P(\Lambda_{\rm ph})$ peaks (diverges) at $\Lambda_{\rm ph}=0$, a small $\Lambda_{\rm ph}$ is statistically preferred. This can be the case if
$P(\Lambda_0)$ peaks (diverges) at $\Lambda_0=0$ and loop and/or string corrections do not significantly modify this peaking behavior.

As we shall discuss in the context of an illustrative $\phi^3/\phi^4$ model, where there is no uncoupled sector and all couplings/parameters are treated as if they are flux parameters so they will take random values within some reasonable ranges, the physical (loop corrected) $V(n_i, \phi_j)$ yields $P(\Lambda_{\rm ph})$ for the physical $\Lambda_{\rm ph}$ while the tree (or bare) $V(n_i, \phi_j)$ yields $P(\Lambda_0)$ for the tree $\Lambda_0$. We find that $P(\Lambda_{\rm ph})$ hardly differs from $P(\Lambda_0)$. Both $P(\Lambda)$s peak (i.e., diverge) at $\Lambda=0$, and the two sets of statistical preferred flux values for $\Lambda \sim 0$ are in general only slightly different. In fact, up to two-loops, $P(\Lambda_{\rm ph})$ is essentially identical to the tree $P(\Lambda_0)$. As a result, although radiative instability may be present, the statistical preference approach actually evades or bypasses this radiative instability problem.
We like to convince the readers that this phenomenon of bypassing the radiative instability problem stays true in more complicated models, as well as when applied to very light scalar boson masses (if present). We also point out how the phase transition issue is circumvented.
 
$\bullet$ The other dS vacuum constructions (KKLT, large volume etc) involve parameters that in principle are calculable but in practice remain unknown and so are treated as arbitrary free parameters. So it remains to be seen whether the conjecture holds in those constructions as well. There are also unknown parameters in the K\"ahler uplift scenario, but the peaking behavior of $P(\Lambda)$ at $\Lambda=0$ turns out to be insensitive to them. 

\begin{figure}[t]
 \begin{center}
 \label{cartoon}
  \includegraphics[scale=0.8]{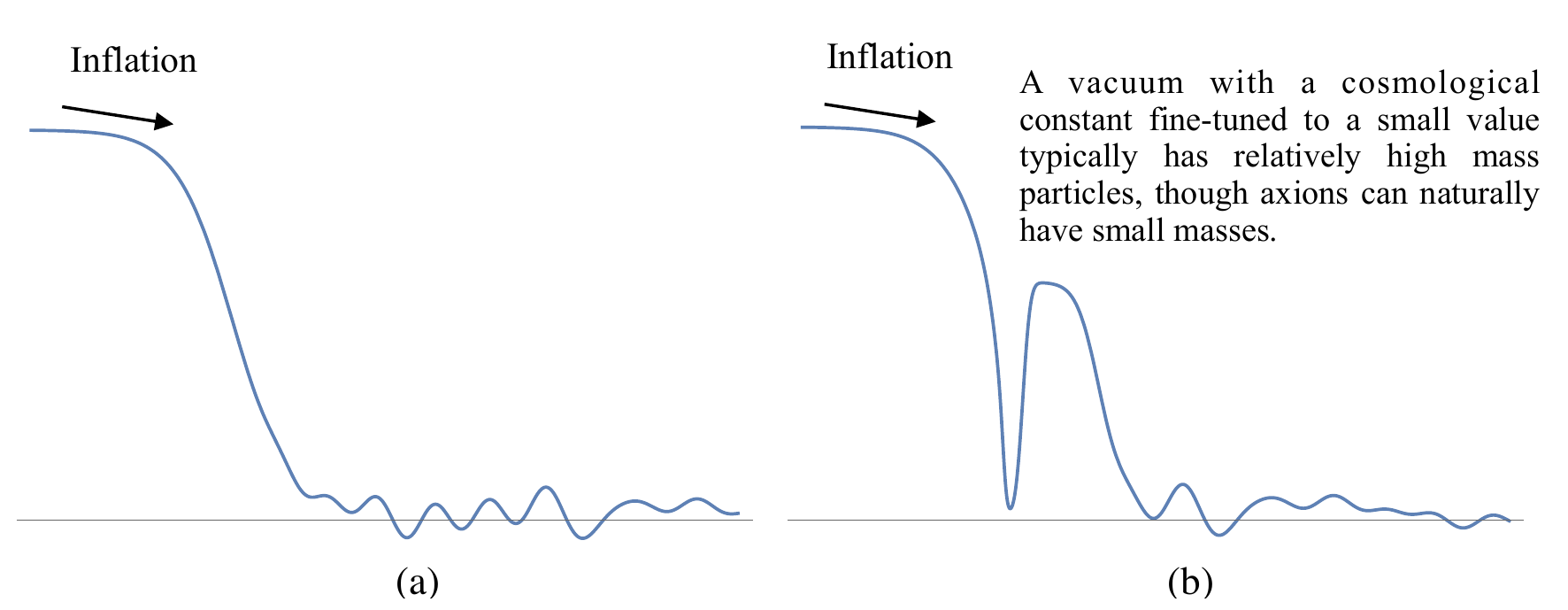}
   \end{center}
    \caption{\footnotesize The left cartoon picture (a) shows the situation where most vacua have a very small cosmological constant; so, after inflation, the universe will roll down to such a low dS vacuum before it has a chance to go to any of the AdS vacua. We argue by examples that such a vacuum has very light bosons. The right cartoon picture (b) shows that if we are allowed to fine-tune free parameters (or by accident), we can also have a vacuum with a very small cosmological constant, so our universe rolls into it. In this case, we typically will get large scalar masses. }
   \end{figure}

$\bullet$ Lest one may think the accumulation of $\Lambda \simeq 0^+$ is due to energetics (i.e., small positive $\Lambda$s are energetically preferred over not so small positive $\Lambda$s), we note that the same accumulation happens for AdS vacua as well; that is, $P(\Lambda)$ peaks (diverges) as $\Lambda \rightarrow 0^-$. 
In fact, typical AdS solutions of $V(n_i, \phi_j)$ involve 2 branches: supersymmetric vacua and non-supersymmetric vacua, where the latter set mirrors the dS solutions (see e.g., \cite{Sumitomo:2012cf}). So for a given range of small $|\Lambda|$, we expect more AdS vacua than dS vacua; that happens even before we relax the constraint to allow light tachyons which do not destabilize the AdS vacua. 
However, there are the following situations to consider for negative $\Lambda$ : 

(1) Let there be a vast number of small $\Lambda$ dS vacua in the cosmic landscape (FIG. \ref{cartoon}(a)). Our universe rolling down the landscape after inflation is unlikely to be trapped by a relatively high dS vacuum, since there is hardly any around. However, since it has to pass through the positive $\Lambda$ region first, it is likely to be trapped at a small positive $\Lambda$ vacuum (as there are many of them) before reaching any AdS vacua (as illustrated in FIG 1(a)). 

(2)  Not every choice of flux parameters in a given model yields a meta-stable vacuum. In such cases, the universe will continue to roll down in the negative $\Lambda$ region, reaching a point where the particular low energy effective potential is no longer valid, especially when one or more moduli attain values larger than $M_S$.

(3) Even if an AdS vacuum is stable against perturbing a modulus $\phi$, it may be non-linearly unstable against some other perturbations \cite{Coleman:1980aw,Bizon:2011gg,Dias:2011ss}.  
This leads us to believe that rolling into an AdS region with a non-zero time-derivative ${\dot \phi}$ and a changing $\phi$ will likely destabilize the classical AdS vacuum.

To avoid issues concerning AdS vacua, we shall focus on dS vacua in this paper for the purpose of phenomenology.  
So in this paper, we normalize $P(\Lambda)$ via ${\int}_{\Lambda \ge 0} P(\Lambda) d \Lambda =1$.

In the construction of dS vacua in the K\"ahler uplift scenario \cite{Rummel:2011cd,Sumitomo:2012vx}, it also becomes clear that an exponentially small $\Lambda$ is invariably accompanied by  exponentially light bosons, i.e., light moduli and their axionic partners. That is, in contrast to a vacuum whose $\Lambda$ is fine-tuned to a very small value (see FIG \ref{cartoon}(b)), we conjecture that 
 
 {\bf A dS vacuum with a naturally small $\Lambda$ tends to be accompanied by very light bosons.}
 
This is not too surprising. Consider the 4-dimensional effective action
\begin{equation}
  \label{Action0}
S=\int dx^4 \sqrt{-g}\left[-\Lambda +\frac{M_P^2}{16 \pi}R - \frac{m_H^2}{2}\Phi_H^2 +  . . .\right]
 \end{equation}
where we have displayed all the relevant operators that are known to be present in nature. If we ignore the $\Lambda$ (the most relevant operator) term, then we have two scales, $M_P \gg m_H$. Why the Higgs mass $m_H$ is so much smaller than the Planck mass $M_P$ poses the well-known mass hierarchy problem. Now knowing that a very small $\Lambda$ is present in nature, we like to know its origin. If its value arises via fine-tuning (or accidentally, see FIG \ref{cartoon}(b)), we have to consider $M_P$ as more fundamental and so are led back to the original mass hierarchy problem.  However, if the smallness of $\Lambda$ arises naturally, in that most of the de Sitter vacua in string theory tend to have a very small $\Lambda$, we should expect scalar masses comparable to the $\Lambda$ scale, as is the case in the models examined.  Following this viewpoint, we may instead wonder why the Higgs mass is so much bigger than $\Lambda$, i.e., $m^2_H \gg \Lambda/M_P^2$. Surely, we should re-examine the mass hierarchy problem in this new light. 

Along this direction, we show that the following scenario can easily happen : the physical mass-squared probability distribution $P_j(m_j^2)$ for some scalar field $\phi_j$ may be peaked at $m_j^2=0$ but the peaking is less strong than that for
 $\Lambda$. If the Higgs boson is such a particle, i.e., $\Phi_H=\phi_j$, then it is natural for 
  \begin{equation}
 \Lambda/M_P^2 \ll m_H^2 \ll M_S^2 \ll M_P^2 .
 \label{HiggsSPA}
  \end{equation}
This statistical preference approach allows us to circumvent the original mass hierarchy problem; that is, a small Higgs mass is natural, not just technically natural. 

One may be concerned that the presence of very light scalars are at odds with observations. However, beyond the weakly interacting massive particle scenario for dark matter, recent study of galaxy formation has led to a renewed interest for a very light boson as the dark matter, with mass $m \simeq 10^{-22}$eV $\simeq10^{-50} M_P$ \cite{Turner:1983he,Hu:2000ke,Goodman:2000tg,Peebles:2000yy,Amendola:2005ad,Boehmer:2007um,Schive:2014dra,Guth:2014hsa,Hui:2016ltb} and with very weak self-couplings \cite{Fan:2016rda}, 
 \begin{equation}
 \label{dark1}
\frac{8 \pi}{3}\frac{m^2}{H^2} \sim  \frac{m^2M_P^2}{\Lambda} \sim 10^{22},
 \end{equation}
where $H$ is the Hubble parameter. Here we explore, within the context of the K\"ahler uplift scenario in string theory that has a naturally small $\Lambda$, the mass spectrum of the light scalars. We see that boson masses in this range is entirely possible within this context. 
 
The rest of the paper is organized as follows. Sec. 2 reviews and extends the $\phi^3/\phi^4$ model discussed in Ref\cite{Sumitomo:2012vx} that captures many (but certainly not all) of the key features that appear in the more elaborate K\"ahler uplift string model which is our main focus. In particular, we review how, in the tree-level version of this $\phi^3/\phi^4$ model, the properly normalized $P(\Lambda)$ peaks (i.e., diverges) at $\Lambda_{tree}=0$. We then discuss the effect of the quantum loop corrections and argue how loop corrections maintain the peaking behavior of $P(\Lambda)$ at the physical $\Lambda_{\rm ph}=0$. Numerically, we see that both one-loop and two-loop corrections have almost no effect on the peaking of $P(\Lambda)$ at $\Lambda=0$. In this sense, the smallness of the physical $\Lambda_{\rm ph}$ is natural, not only technically natural. We point out that radiative instability may be present in this model, and explain how the statistical preference approach actually bypasses this radiative instability problem. 
We also find that $P(m^2)$ for the $\phi$ boson mass does not peak at $m^2=0$. Since the peaking of $P(\Lambda)$ at $\Lambda=0$ is very weak here, and we do expect that the peaking of $P(m^2)$, if any, to be weaker than that for $P(\Lambda)$, so the non-peaking of $P(m^2)$ at $m^2=0$ in this model is consistent with our picture.

Sec. 3 reviews the calculation of $\Lambda$ and its probability distribution in a K\"ahler uplift model within flux compactification in Type IIB/F theory. This simplified yet non-trivial model, with an arbitrary number $h^{2,1}$ of complex structure moduli, is first studied in Ref\cite{Rummel:2011cd} while the probability distribution $P(\Lambda)$ of $\Lambda$ is discussed in Ref\cite{Sumitomo:2012vx}.  We choose this model partly because it can be solved semi-analytically. The model is first solved for a supersymmetric AdS solution which is then K\"ahler uplifted to a dS vacuum. The K\"ahler uplift in this model relies on the known perturbative $\alpha'^3$ correction and a non-perturbative term for the K\"ahler modulus. We see that the median of $\Lambda$ can be as small as the observed value (\ref{L1}) if $h^{2,1} \sim {\cal O}(100)$.

Sec. 4 determines the scalar mass spectrum when $\Lambda$ is very small.  Some preliminary studies on this issue can be found in Ref\cite{Rummel:2011cd,Sumitomo:2012vx}. It is shown there that K\"ahler uplift will shift the boson masses by relatively small amounts, i.e., ${\delta m^2}/m^2$ is suppressed by powers of the (dimensionless) compactification volume $\cal V$ and/or powers of $h^{2,1}$. Since the string scale $M_S$ is around the GUT scale, the compactification volume ${\cal V} \sim {\cal O}(10^3)$ and $h^{2,1} \sim {\cal O}(100)$, we shall first find the boson mass spectrum coming from  the AdS solution before uplifting to a dS vacuum. This approximation is in fact good enough for our purpose. Including the dilaton (but not the K\"ahler modulus), we have $h^{2,1}+1$ complex bosons. The mass matrix for the scalar ones decouples from that for the pseudo-scalar ones (axions). Diagonalizing them yields:

$\bullet$ ($h^{2,1}-2$) of the pseudo-scalars stay massless. These axions are expected to gain masses via  
non-perturbative instanton effects.

$\bullet$ ($h^{2,1}-2$) scalars have a degenerate mass twice the gravitino mass $m^2 =  4e^KW^2$.

$\bullet$ Three in each set obtain heavier masses, where the heavier ones can have masses in the range (\ref{dark1}) as potential candidates for light dark matter.

$\bullet$ The K\"ahler modulus has a massless axion and a scalar mass comparable to the other scalar masses. Again, the axion is expected to gain a small mass via instanton effect.

$\bullet$ Couplings among themselves are also extremely weak.
 
$\bullet$ Some of the very light bosons can be made heavier by turning on non-geometric fluxes.

 Since the string theory scenarios studied here are simplified versions of actual flux compactifications and still far from particle physics phenomenology, the discussion is limited to generic orders-of-magnitude features only.  

Sec. 5 discusses the moduli masses in a Racetrack K\"ahler uplift scenario. After a brief review on how $\Lambda$ can be exponentially small here, we also point out that the scalar masses are exponentially small, just like the value of $\Lambda$. Here we see how an axion with a small mass can have a small repulsive self-interacting term. Some discussions are put in Sec. 6, including a brief discussion on the cosmological production of these light bosons.
Sec. 7 presents some remarks and our conclusion. Some details have been relegated to the appendix.

\section{An Illustrative $\phi^3/\phi^4$ Toy Model}

The statistical preference for a small $\Lambda$ follows if the low energy effective potential has no continuous free parameter and all sectors are connected via interactions, as is the case in string theory; that is, it is a function of only scalar fields or moduli, quantized flux values, discrete values like topological indices, and calculable quantities like loop and string corrections, with no disconnected sectors.   To get some feeling on some of these features, let us review the single scalar field polynomial model discussed in Ref\cite{Sumitomo:2012vx}. In this model, gravity and so $M_P$ is absent. So the statistical preference for a small $\Lambda$ shows up only as the (properly normalized) probability distribution $P(\Lambda)$ peaks at $\Lambda=0$, in particular when $P(\Lambda)$ diverges there, i.e., 
  \begin{equation}\label{phipeak}
\lim_{\Lambda \rightarrow 0^+} P(\Lambda) \rightarrow \infty.
 \end{equation}
The divergence of $P(\Lambda=0)$ is rather mild here, so it is far from enough to explain the very small observed value of $\Lambda$ (\ref{L1}); but it does allow us to explain a few properties that are relevant for later discussions. As the number of moduli and flux parameters (coupled together) increases, we do expect the divergence of $P(\Lambda=0)$ to be much sharper (so the median of $|\Lambda|$ decreases), as illustrated by the string theory models.

Consider the tree level potential,
  \begin{equation}
  \label{A11}
V_0(\phi) = a \phi - {b \over 2} \phi^2 + {c \over 3!} \phi^3 +\frac{d}{4!}\phi^4,
 \end{equation}
 where $\phi$ is a real scalar field, mimicking a modulus. We are not allowed to introduce a ``constant" or flux parameter  term by itself since it will be disconnected to the $\phi$ terms in $V_0(\phi)$. 
Imposing the constraint that the tree level  $V_0$ has no continuous free parameter except some scale $M_s$, the parameters $a$, $b$, $c$ and $d$ mimic the flux parameters that take only discrete values of order of the $M_s$ scale, thus spanning a ``mini-landscape". Let them take only real values for simplicity. We may also choose units so $M_s=1$. For a dense enough discretuum of $\Lambda$, a flux parameter may be treated as a random variable with continuous value over some range. Let us look for dS solutions with flux parameters $a$, $b$, $c$, 
$d\in [0,1]$ or some other reasonable range. We start with the tree-level properties, where (\ref{phipeak}) is satisfied, and then discuss the multi-loop corrections. We argue that the peaking behavior (\ref{phipeak}) remains present when we include multi-loop corrections, that is, when $P(\Lambda)$ is for the physical $\Lambda_{\rm ph}$. We also explain how the statistical preference approach bypasses the radiative instability problem even if it is present.

Starting with the tree-level effective potential $V_0(\phi)$ (\ref{A11}), we impose the stability $M^2=\partial_\phi^2 V_0|_{v_0} >0$ at the extremal points given by $\partial_\phi V_0{\big |}_{v_0}=0$, with each vacuum expectation value $v_0$ yielding $\Lambda_0(v_0) = V_0(v_0)$ and
  \begin{equation}
  \label{A22}
  M_0^2=\frac{\partial^2 V_0}{\partial \phi^2}{\big |}_{v_0} = -b+cv_0 + dv_0^2/2, \quad \quad  \lambda =\frac{\partial M^2}{\partial v}=c+dv_0.
 \end{equation}
We study three case : the $\phi^3$ model with $c=1$ and with random $c$, and the $\phi^4$ case with random flux parameters $\{a, b, c, d\}$.

\subsection{The $\phi^3$ Model at Tree-Level with $c=1$}

\begin{figure}
\begin{center}
  \includegraphics[scale=0.6]{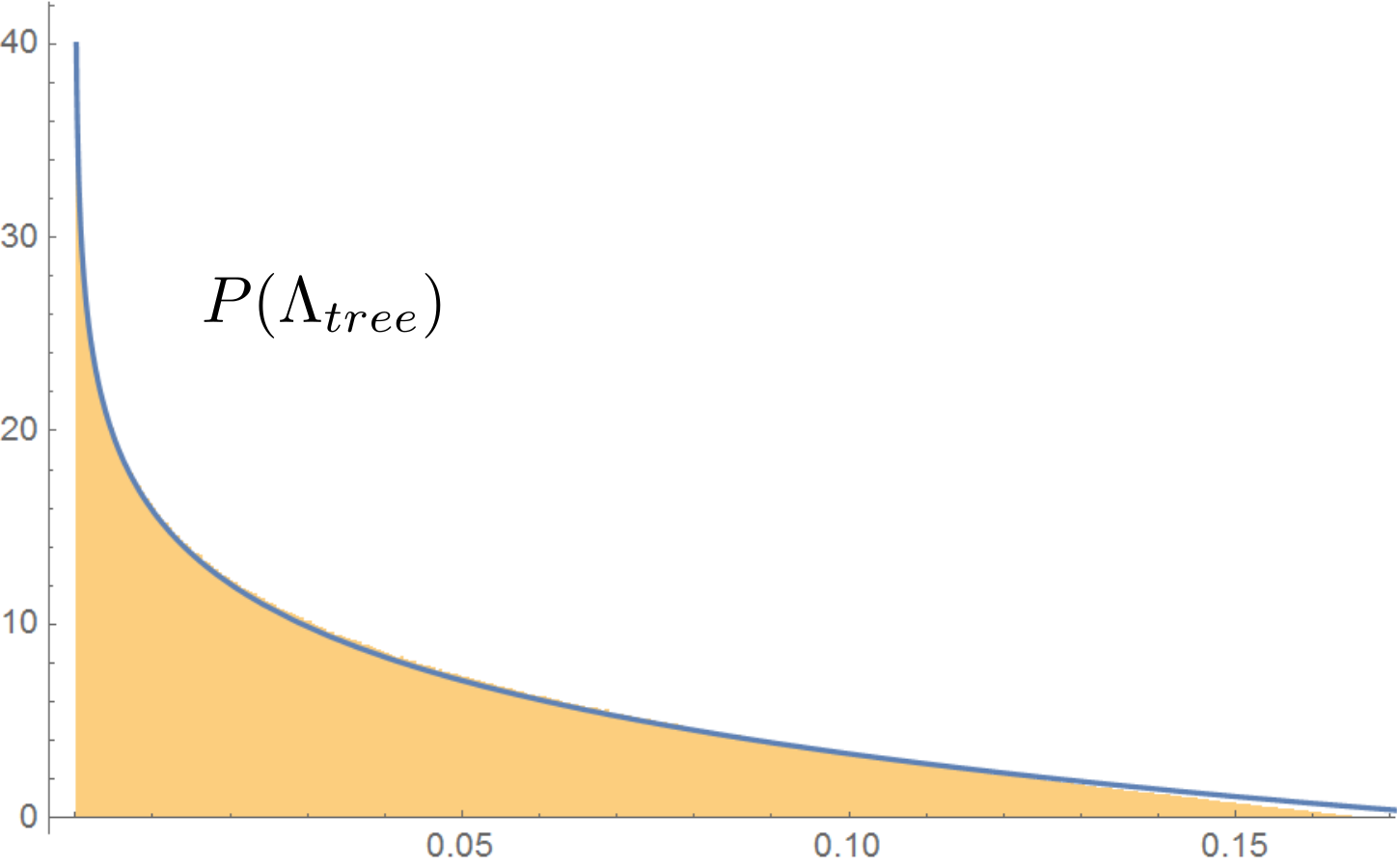}
  \includegraphics[scale=0.6]{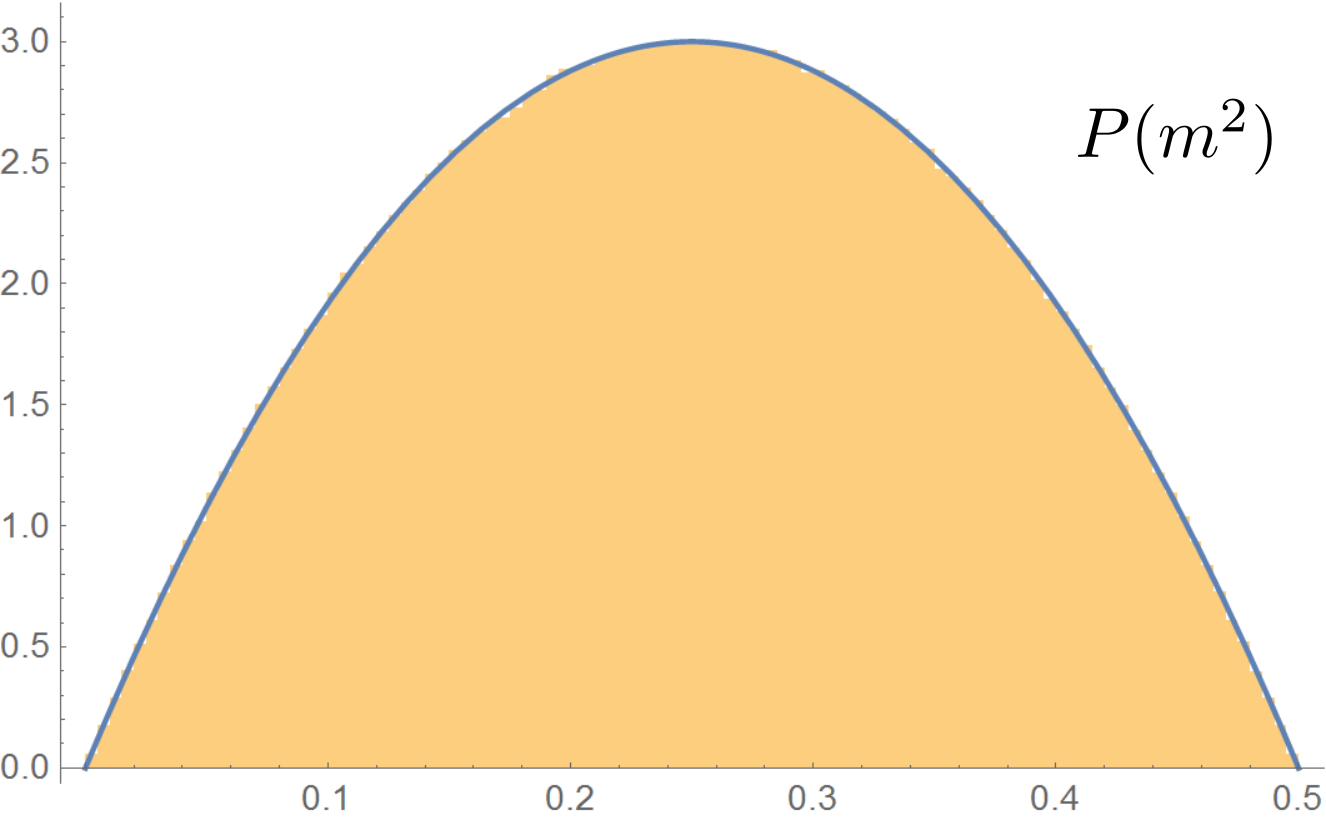}
  \end{center}
  \caption{Probability distribution $P(\Lambda_{0})$ (left) and $P(m^2)$ (right) for the tree-level $\phi^3$ model with coefficient $c=1$. Histograms under the (approximate) analytic curves are numerical data. Here $P( {\Lambda_0})$ diverges logarithmically at  $\Lambda_0=0$, while $P(m^2)$ peaks at $m^2 \sim 0.26$.}\label{c1lm}
\end{figure}

At least a polynomial of degree three (with no constant term) is required for a metastable vacuum with a positive $\Lambda$, so let us start with
$d=0$.
Requiring the stability $\partial_\phi^2 V_0{\big |}_{min} >0$ at the extreme points given by $\partial_\phi V_0=0$ yields
\begin{equation}
 \Delta \equiv \sqrt{b^2-2 ac} >0, \quad \phi_{0,min}= (b + \Delta)/{c}
 \label{Delta}
\end{equation}
\begin{equation}
\Lambda_0\equiv V_{0,min}= \frac{(b+\Delta)^2(b-2\Delta)}{6 c^2}.
\label{A12}
\end{equation}
Taking smooth distributions $P(a)$, $P(b)$ and $P(c)$ when $a$, $b$ and $c$ take dense discrete values, one finds that the probability distribution $P( {\Lambda_0})$ of positive $\Lambda_0$ diverges as $\Lambda_0 \rightarrow 0^+$. This peaking of $P( {\Lambda_0})$ at $\Lambda_0=0$ also happens if we fix $c=1$, so let us consider this case in more detail.
Now, using Eq.(\ref{Delta}) and Eq.(\ref{A12}), we have kinematical constraints. 
\begin{equation}
1  \ge b^2 \ge \Delta^2 + 3/4, \quad \quad 1/2 \ge \Delta \ge 0 \rightarrow \Lambda_0 \le 1/6.
\end{equation}
Performing a change of variable to $\delta$,
\begin{equation}
 P(\Lambda) = N \int d\Delta \int_0^1 d {a} d {b} \, \delta\left( \Lambda -(b+\Delta)^2(b-2\Delta)/6 \right)\, \delta\left(\Delta -\sqrt{b^2-2a} \right).
\end{equation}
Integrating over the 2 $\delta$-functions, we obtain, 
\begin{equation}
P(\Lambda) \propto \int  \frac{\Delta}{(b^2-\Delta^2)} d\Delta \propto \ln (b^2-\Delta^2).
\label{delta3}
\end{equation}
Since $P(\Lambda) \rightarrow \infty$ as  $\Lambda  \rightarrow 0$,
$P(\Lambda)$ diverges logarithmically, as it should be. (One can also use $\int_D  \delta(g(x)) d^n x = \int_{g^{-1}(0)}|\nabla g|^{-1} d^{n-1}\sigma $ to verify that $P(\Lambda=0) \to \infty$.) 
One then finds a formula for $P( {\Lambda_0})$ 
\begin{equation}
 P( {\Lambda_0}) \simeq \frac{16}{3} \ln \left(\frac{3}{16\Lambda_0}\right).
  \label{eq:divergent PDF}
\end{equation}
That is, $P(\Lambda_0)$ is divergent at ${\Lambda_0} = 0$, as shown in FIG \ref{c1lm}. This divergence remains even if $c$ takes dense discrete values as well (see FIG \ref{randomc}). Although this logarithmic peaking behavior (and so the statistical preference for $\Lambda_0=0$) is very weak, it does show that $\Lambda_0=0$ is special.
We also find that the probability distribution of the mass squared of $\phi$, $m^2=\Delta=\sqrt{b^2-2a}$
\begin{equation}
   P(m^2) = \int da db \delta(m^2-\sqrt{b^2-2a}) = 24(1-2m^2)m^2, \quad  \text{or} \quad P(m^4) = 12(1-2\sqrt{m^4}).
\end{equation}
So, in this case, there is no statistical preference for a massless mode or a very light $\phi$.

To summarize some of the lessons learned in this simple model : \\
$\bullet$ Not all choices of flux values for $\{a, b, c\}$ will yield a classically stable vacuum. \\
$\bullet$ Adding an arbitrary constant to $V_0(\phi)$ will surely remove the preference for $\Lambda_0=0$. However, in string theory, there is no such arbitrary constant we can include, since string theory has no free continuous parameter (besides $M_S$ that sets the string scale). Furthermore, all fields and fluxes are coupled via the closed string sector, so there is no uncoupled sectors.  \\
$\bullet$ The peaking of $P(\Lambda_0)$ at $\Lambda_0=0$ is insensitive to the input probability distributions for the flux parameters as long as they are smooth enough with large enough ranges. If $a=b \neq c$, the logarithmic peaking (\ref{eq:divergent PDF}) strengthens to $P(\Lambda_0) \simeq 1/\sqrt{\Lambda_0}$ for $\Lambda_0 \gtrsim 0$. On the other hand, the peaking disappears if $a=c \neq b$, $b=c \neq a$ or $a=b=c$. If $a \sim n_a^2$ while $P(n_a)$ for the discrete flux value $n_a$ is smooth around $n_a=0$, and/or similarly for $b$, then $P(\Lambda_0)$ is more sharply peaked at $\Lambda_0=0$ than that given in Eq.(\ref{eq:divergent PDF}). One can also choose $P(a)$ and $P(b)$ so that $P(m^2)$ also peaks at $m^2=0$. \\
$\bullet$ As we shall see in FIG \ref{randomc}, adding a $d\phi^4/4!$ term to $V_0(\phi)$ (where $d$ takes discrete flux values) does not change the qualitative peaking behavior of $P(\Lambda_0)$ at $\Lambda_0^+=0$, which is also maintained if we add higher powers of $\phi$, say $\phi^6$, to the potential $V(\phi)$ (\ref{A11}). \\
$\bullet$ If we include $\Lambda_0 \le 0$ solutions, we find that $P(\Lambda_0)$ also peaks at $\Lambda_0=0^{-}$. \\
$\bullet$ For other toy models on small $\Lambda$, see Ref\cite{Brown:2013fba,Brown:2014sba}. \\

\subsection{Loop Corrections}

It is convenient to calculate the multi-loop contributions to the tadpole diagrams using the dimensional regularization method and then integrate them to obtain the effective potential $V(\phi)$. We see that the $n$-th loop contribution to $V(\phi)$, namely $V_n(\phi)$, is a function of $M^2(\phi)$, $\lambda(\phi)$ (\ref{A22}), and $d$ only. At the one-loop level, $V_1(\phi)$ is a function of $M^2(\phi)$ only. 
Simple dimensional reasoning yields
  \begin{align}
 V(\phi) = V_0+ \sum_{n \ge1} V_n = M^4F\left(\frac{\lambda^2}{M^2}, \ln ({M^2}), d\right),\nonumber \\ M^2(\phi)=V''_0(\phi),\quad \lambda(\phi)=V'''_0(\phi), \quad d(\phi)=V''''_0(\phi),
   \end{align}
where each prime stands for a derivative with respect to $\phi$ and $F(\lambda^2/M^2, \ln(M^2), d)$ is a polynomial in the dimensionless parameters $\lambda^2/M^2, \ln(M^2)$ and $d$. More precisely, for $n \ge 1$,
$$V_n = \frac{M^4}{(4 \pi)^{2n}} f_n \left({\lambda^2}/{M^2}, \ln ({M^2}), d\right), $$
where $f_n$ is a polynomial up to $n$-th power in $\ln (M^2)$, and $(n-1)$-th (combined) power in ${\lambda^2}/{M^2}$ and $d$, with $n$-dependent coefficients which grow much slower than the $(4 \pi)^{2n}$ factor. 

The peaking behavior (\ref{phipeak}) for $\Lambda$ remains if the probability of the loop correction size
$P({\sum V_n}/{\Lambda_0})$ is suppressed for $|\sum V_n| > \Lambda_0 \sim 0^+$. That is, only a small fraction of the small $\Lambda$ cases are impacted, so the majority of the $ \Lambda_0 \simeq 0^+$ cases remains to contribute to the peaking of $P( \Lambda)$. Considering individual terms in $f_n \left({\lambda^2}/{M^2}, \ln ({M^2}), d\right)$, we see that this is easily satisfied if the coefficients of the terms in $f_n \left({\lambda^2}/{M^2}, \ln ({M^2}), d\right)$ grow no faster than a small positive power of $n$. For example,
we see numerically that $P({M^4 \ln M^2}/{64 \pi^2\Lambda_0})$ versus $\log[ |M^4 \ln M^2|/{64 \pi^2\Lambda_0}]$ has an approximate Gaussian distribution that peaks at a few percent of $|M^4 \ln M^2| /{64 \pi^2\Lambda_0}$ for small $\Lambda$, i.e., it is heavily suppressed for $|M^4 \ln M^2| >{64 \pi^2\Lambda_0}$. This means that the peaking behavior of $P(\Lambda)$ is at most slightly modified. In short, we see that the peaking behavior (\ref{phipeak}) for $\Lambda$ remains if
\begin{equation}
\label{rscond}
\lim_{\Lambda_0 \simeq 0^+} |V_n| < \Lambda_0.
\end{equation}
In other words, the loop corrections converge in a way that, for most choices of flux parameters (but not all), the loop corrected $\Lambda$ does not differ much from the tree $\Lambda_0$. That is, $P(\Lambda_0)$ peaks at the tree $\Lambda_0=0$ and $P(\Lambda_{\rm ph})$  peaks at the physical $\Lambda_{\rm ph}=0$.  It is easy to see numerically that this is true. 
Naturalness of the smallness of $\Lambda$ implies it is technically natural as well (but not the other way).
As an illustration, let us show the peaking behavior of $P(\Lambda)$ at $\Lambda \sim 0$ for the $\phi^3$ and the $\phi^4$ models up to 2-loops.  We then address the radiative instability issue, which appears when the loop-corrected $\Lambda$ differs substantially from the tree $\Lambda$.

\subsection{The One-Loop and Two-loop Cases} 

\begin{figure}[h]
\begin{center}
\includegraphics[scale=0.65]{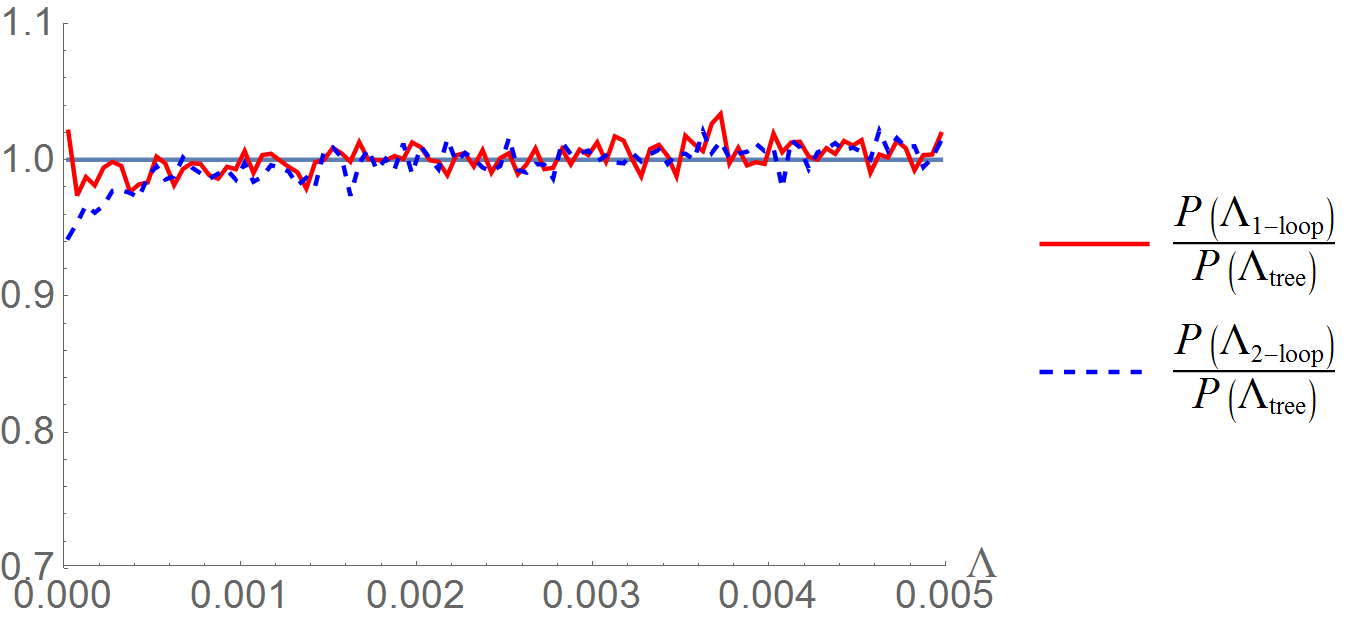}
\end{center}
\caption{The probability distributions $P(\Lambda)$ for the $\phi^3$ ($c=1$) model for tree $P(\Lambda_{0})$, one-loop corrected $P(\Lambda_1)$ and two-loop corrected $P(\Lambda_{2})$.  Shown are the ratios between the distributions (numerical), ${P(\Lambda_1)}/{P(\Lambda_{0})}$ (red solid) and ${P(\Lambda_{2})}/{P(\Lambda_{0})}$ (blue dash) at small values of $\Lambda$. The horizontal solid line at 1.0 stands for $P(\Lambda_{0})$, which  is shown in FIG \ref{c1lm}.} \label{c1p3}
\end{figure}

The key of a naturally small $\Lambda_{\rm ph}$ depends on its functional dependence on the flux values, which is different from that for $\Lambda_0$. Here we consider the explicit forms of the one- and two-loop corrections to $\Lambda$.
First, let us introduce the one-loop radiative correction \cite{Coleman:1973jx,Weinberg:1973ua} to the tree-level potential $V_0(\phi)$ (\ref{A11}),
\begin{equation}
\label{toy1}
\begin{split}
 V(\phi) = & V_0(\phi) +V_1(\phi) = V_0(\phi) +  \frac{1}{64 \pi^2} M^4 \left(\ln (M^2) -{1\over 2}\right) .
\end{split}
\end{equation}
Now, the minimum of $V(\phi)$ is shifted from $v_0$ to $v_1=v_0 + \delta v$, so the physical 
$\Lambda_{\rm ph} = V_0(v_1)+V_1(v_1)$,
where, to leading order,
\begin{equation}
\begin{split}
\delta v &= -V_1'(v_0)/M_0^2 = -\frac{\lambda_0}{32\pi^2} \ln M_0^2, \\
\Lambda_{\rm ph} &= \Lambda_1= \Lambda_0 +V_1(v_0) - \frac{1}{2} \frac{(V_1'(v_0))^2}{M_0^2}.
\end{split}
\end{equation}
In the $\phi^3$ case, $M_0^2=\Delta= -b+c\phi_{0,min}$ (\ref{Delta}).
On one hand, the one-loop contribution shifts $\Lambda$ to a smaller value, so some of the $\Lambda_0 \simeq 0$ cases have been shifted to negative $\Lambda$s, depleting the peaking of $P(\Lambda)$ at $\Lambda=0$. On the other hand, the flux parameter region contributing to small $\Lambda$ region grows, enough to compensate for the loss. This can be seen as $b^2 \ge 2ac \rightarrow b^2 \ge 2ac - |\delta v| M_0^2$, thus enlarging the region of parameter space contributing to $\Lambda \simeq 0^+$. It is this region that provides additional contributions to the peaking of $P(\Lambda)$ at $\Lambda=0^+$. As a result, the peaking for the one-loop corrected $P(\Lambda)$ is comparable to that for the tree-level $P(\Lambda)$, as shown in FIG \ref{c1p3}.

\begin{figure}
\begin{center}
\includegraphics[scale=0.4]{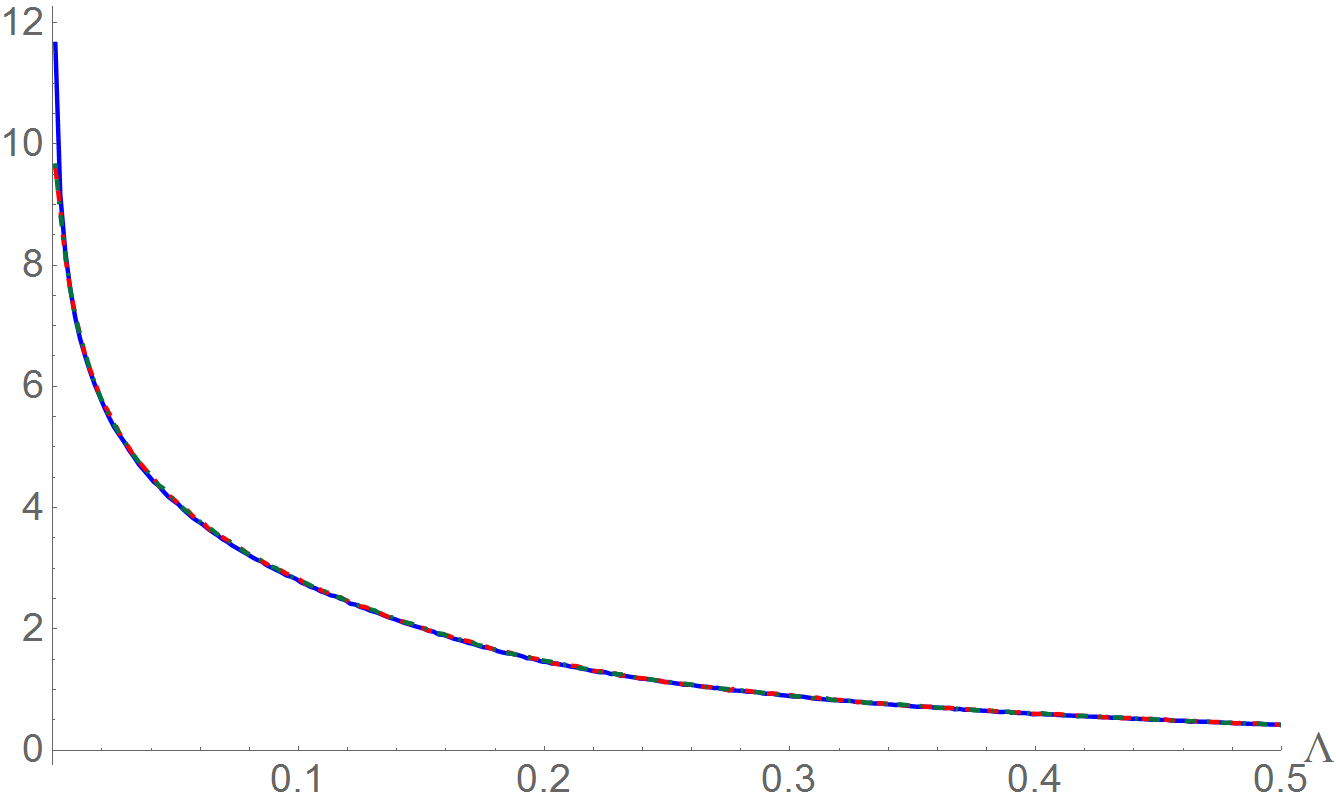}
\includegraphics[scale=0.5]{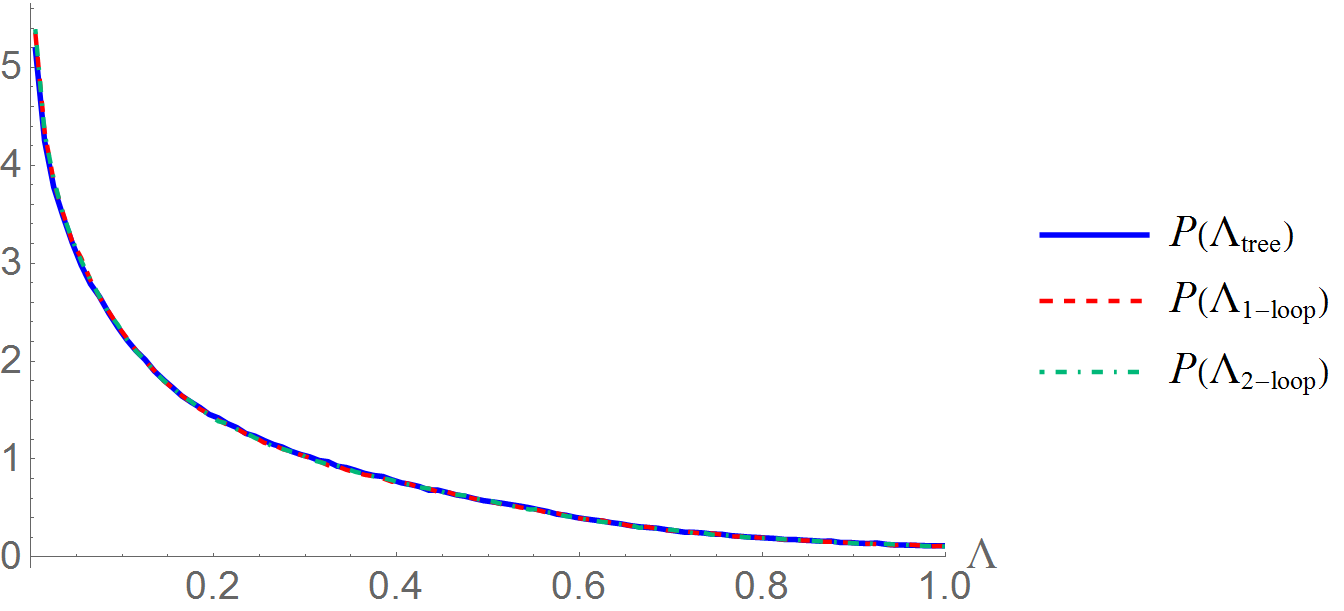}
\end{center}
\caption{Probability distribution $P(\Lambda)$ for the $\phi^3$ model (left) with $c$ randomized in $[0,1]$; and for the $\phi^4$ model given in Eq(\ref{A11}) and Eq(\ref{phi4d}) (right). In each case, the blue solid curve is for the tree-level $P(\Lambda_0)$, the red dashed curve is for the one-loop corrected $P(\Lambda_1)$ and the green dot-dash curve is for the two-loop corrected $P(\Lambda_2)$. In each case, the loop-corrected and the tree $P(\Lambda)$s are essentially on top of each other, showing that loop corrections have little impact on the distribution $P(\Lambda)$. In particular, the peaking behavior of $P(\Lambda)$ at $\Lambda=0$ remains intact.
} \label{randomc}
\end{figure}

The two-loop correction is given by \cite{Jackiw:1974cv,Lee:1974fj}
\begin{equation}
V_2(\lambda, M^2, d)=  \frac{M^4}{(4 \pi)^4}\left( \frac{\lambda^2}{M^2} \bigg[\frac{1}{4} \ln^2M^2-\frac{9}{7}(\ln M^2-1) \bigg] + d \bigg[\frac{1}{4}\ln^2 M^2 + \frac{1}{4}\ln M^2 -\frac{79}{28} \bigg] \right)
\end{equation}
so the two-loop renormalized minimum is now shifted to $v_2$, i.e.,  $\frac{\partial V(\phi)}{\partial \phi}\big|_{v_2}=0$, and the two-loop renormalized $\Lambda_2$ is given by
$$\Lambda_2=V(v_2)=V_0(v_2) +V_1(v_2) +V_2(v_2).$$
Going back to the $\phi^3$ model with $c=1$, we find that the loop corrected $P(\Lambda_1)$ and $P(\Lambda_2)$ are very close to the tree $P(\Lambda_{0})$ shown in FIG \ref{c1lm}. FIG \ref{c1p3}
shows the ratio of the probability distributions ${P(\Lambda_1)}/{P(\Lambda_{0})}$ and ${P(\Lambda_{2})}/{P(\Lambda_{0})}$ for small values of $\Lambda$. 
At least up to two-loops, $P(\Lambda_{\rm ph})$ continues to peak (diverge) at  $\Lambda_{\rm ph}=0$. The same behavior is true for the $\phi^3$ model with a random $c$. The loop corrected $P(\Lambda_1)$ and $P(\Lambda_2)$ in this case are essentially indistinguishable from the tree $P(\Lambda_{0})$, as shown in FIG \ref{randomc}(left).

Now consider the $\phi^4$ model. Adding a $d \phi^4 /4!$ term does not change the qualitative peaking behavior of $P(\Lambda=0)$. Take the $\phi^4$ potential (\ref{A11}) where the corresponding flux parameter regions are,
\begin{equation}
\label{phi4d}
a\in [-1,1],\quad b \in [-1,1],\quad c \in [-1,1], \quad d\in [0,1].
\end{equation}
As we vary the flux parameters in the above region, we may get none or more than one positive (local) minimum for any specific choice of $\{a, b, c, d\}$. The probability distribution $P(\Lambda)$ for the tree-level $\Lambda_0$, the one-loop renormalized $\Lambda_1$ and the two-loop renormalized $\Lambda_2$ are shown in FIG \ref{randomc} (right), where we present $P(\Lambda)$ for the $\phi^4$ model (\ref{phi4d}). Again, we see that the loop corrected $P(\Lambda_1)$ and $P(\Lambda_2)$ are essentially indistinguishable from the tree $P(\Lambda_{0})$, verifying the statement that loop corrections have negligible effect on the peaking behavior of $P(\Lambda)$ at small $\Lambda$. 

To summarize, the statistical preference for $\Lambda=0$ remains, for either the tree-level $\Lambda_0$ or the loop-corrected $\Lambda_{\rm ph}$. Although the functional dependence of $\Lambda$ on the flux parameters are different for $\Lambda_0$ and $\Lambda_{\rm ph}$,  
nevertheless, given the same probability distributions for the flux parameters, we see that $P(\Lambda_{\rm ph})$ is essentially the same as  $P(\Lambda_0)$.
It will be nice to investigate the above properties for more general quantum field theory models that satisfy the stringy conditions : no free parameters except flux parameters and no uncoupled sectors. Of course, the cases we are really interested in are the flux compactifications in string theory. However, we do gain some intuitive understanding from   examining this relatively simple model. 

In a more realistic model to explain the observed $\Lambda_{obs}$, $P(\Lambda)$ has to diverge at $\Lambda_{\rm ph}=0$ much more sharply than the logarithmical divergence shown in this model.
In more non-trivial models in string theory to be discussed below, we envision that both $P(\Lambda_{\rm ph})$ for $\Lambda_{\rm ph}$ and $P(m^2_{\rm ph})$ for the some bosons prefer small values, while the peaking in $P(\Lambda_{\rm ph})$ can be much stronger than that in $P(m^2_{\rm ph})$.  
If one applies this to the Higgs boson in a phenomenological model, the observed situation (\ref{HiggsSPA}) can follow from their statistical preferences. 

Without showing details, we find that, in this model, $P(m^2)$ does not peak at $m^2=0$ in every case considered above, loop corrected or not, as illustrated by the simple case shown in FIG \ref{c1lm}, although $P(m^4)$ does peak (but does not diverge) at $m^2=0$.  That is, this model shows no sign that a light boson is preferred. Since the peaking of $P(\Lambda)$ is so very weak already, and the preference for small mass squared is expected to be even weaker, this property is consistent with our general qualitative picture. This also means this simple model cannot address the Higgs boson mass hierarchy problem. It will be very interesting to study other quantum field theory models to see whether a light scalar mass will be statistically preferred.

 \subsection{Bypassing the Radiative Instability Problem}

Now we have seen that the statistical preference for small $\Lambda$ is robust. Although the set of flux parameters that yield a small tree-level $\Lambda_0$ and the set of flux parameters that yield a small physical $\Lambda$ largely overlap, they are not identical. For the non-overlapping choices, radiative instability may be present. Here we like to explain how the statistical preference approach simply bypasses this radiative instability problem.

In usual quantum field theory, we can fine-tune the parameters/couplings in the tree-level effective potential  to obtain a very small $\Lambda_0$.  It turns out that the radiative correction typically overwhelms the small tree-level value $\Lambda_0$, so one has to fine-tune the parameters/couplings again to obtain a small $\Lambda_{\rm ph}$. This fine-tuning has to be repeated each time a higher order quantum correction is included.  This phenomenon is known as radiative instability. Let us see how the statistical preference approach bypasses this radiative instability problem. We may simplify the discussion by considering the one-loop $\phi^3$ case and fixing $c=1$ without affecting the qualitative peaking behaviors.

\begin{figure}[h]
 \begin{center}
  \includegraphics[scale=.8]{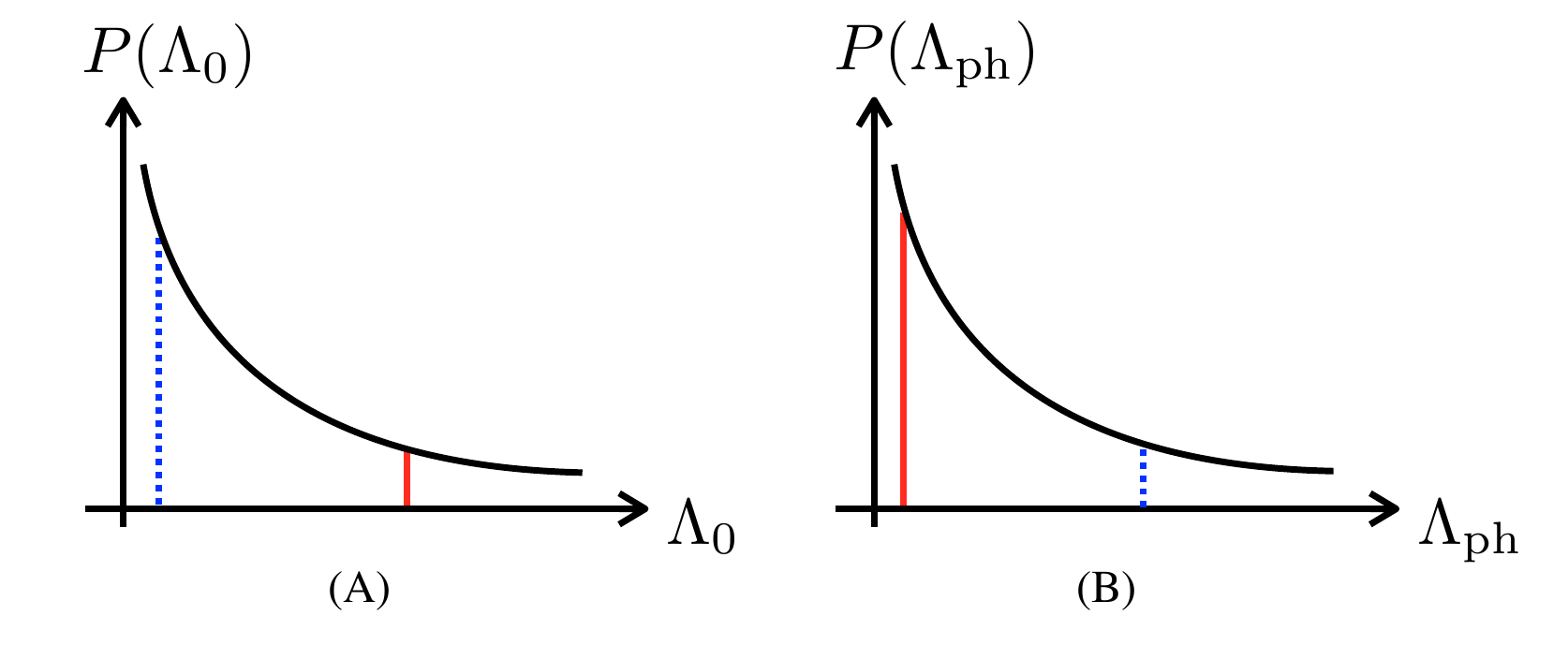}
 \end{center}
 \caption{The black curves in these schematic plots are the normalized probability distributions $P(\Lambda)$ for the tree-level $\Lambda_0$ (A) and for the physical  $\Lambda_{\rm ph}$ (B). Both $P(\Lambda)$ diverge logarithmically at $\Lambda=0$. The red solid lines represent $\Lambda_0$ and $\Lambda_{\rm ph}$ calculated with flux values $(a_1,\, b_1)$ while the blue dotted lines represent $\Lambda_0(a_0,b_0)$ and $\Lambda_{\rm ph}(a_0,b_0)$. Here, the small values $\Lambda_0(a_0,b_0)$ and $\Lambda_{\rm ph}(a_1,\, b_1)$ are statistically preferred.} \label{AToy}
\end{figure}

That  $P(\Lambda)$ peaks (i.e., diverges) at $\Lambda=0$ means there are more vacua with $\Lambda \sim 0$ than vacua with larger $\Lambda$. That is, a random choice of flux values is likely to yield a vacuum with a small $\Lambda$. Suppose we make a random choice of flux values $(a_0, b_0)$. Because of the statistical preference, the resulting vacuum is likely to have a small $\Lambda_0$, as shown schematically in FIG \ref{AToy}(A) (blue dotted line).  Next we introduce the two-loop corrected $\Lambda_2$. Depending on the choice of flux values $a_0$ and $b_0$, there are at least the following 2 possibilities : 
$$\left | \delta \Lambda/\Lambda_0\right | \lesssim 1 \quad {\rm or}  \quad \left | \delta \Lambda/\Lambda_0 \right | \gg 1 $$ 
where $\delta \Lambda=\Lambda_2-\Lambda_0$.
In the first (statistically likely) case, with a small $\Lambda_0$, $\Lambda_{\rm ph}=\Lambda_2$ stays small.  
In the second (statistically less likely) case, radiative correction overwhelms the tree-level value $\Lambda_0$, so $\Lambda_2=\Lambda_{\rm ph}(a_0, b_0)$ ends up relatively big, as shown in FIG \ref{AToy}(B) (the blue dotted line).  
This is radiative instability; though unlikely, it does happen in this $\phi^3$ model. (This qualitative scenario persists if we turn on the $\phi^4$ term and/or render the parameter $c$ random.)
 Let us focus on this second case in which radiative instability happens.
 
For illustration, take for example, the following two choices of flux parameters in $V_0(\phi)$ in the $\phi^3$ model with $c=1$,
\begin{align*}
(a_0,b_0) &= (9.37501\times 10^{-4},1/20), \quad  \Lambda_0 =7.5\times 10 ^{-10}   , \quad \Lambda_{\rm ph} = 5.25\times 10^{-6},    \\
(a_1,b_1) &= (9.39533\times 10^{-2} ,1/2), \quad  \Lambda_0 = 1.52\times 10 ^{-4} , \quad \Lambda_{\rm ph} = 2.5\times 10^{-8} .
\end{align*}
(where, for the sake of discussion, $\Lambda$ is considered to be small if  $\Lambda < 10^{-7}$.)
These two choices are shown schematically in FIG \ref{AToy}, where the first choice gives blue dashed lines while the second choice gives red solid lines.
In the first choice, even though $\Lambda_0$ is small, $\Lambda_{\rm ph}=\Lambda_2$ is not. To obtain a small $\Lambda_{\rm ph}$, we have to start with a different choice of parameters, say the second set $\{a_1,b_1\}$, which yields a tree-level $\Lambda_0(a_1,b_1)$ which may not be small (as indicated schematically by the red line in FIG \ref{AToy}(A)).  
Here, the radiative correction is big enough to bring a not so small $\Lambda_0$ to a small $\Lambda_{\rm ph}$. That is, we have to ``fine-tune"  the parameters in the model to obtain a small $\Lambda_{\rm ph}$. This is the radiative instability problem. It means that, to obtain a small $\Lambda_{\rm ph}$, fine-tuning has to be applied to the couplings/parameters in the field theory model each time we include a higher order radiative correction.

It should be clear how the statistical preference approach bypasses this radiative instability problem. First, we have no parameters to be fine-tuned, since we are already sweeping through all allowed values of the parameters/couplings.  That is, there is no fine-tuning to be done. Instead, we find that the peaking of  $P(\Lambda_{\rm ph})$ at $\Lambda_{\rm ph}=0$ is present, so a ``statistically preferred" $\Lambda_{\rm ph}$ should be small, with some flux values $\{a_1, b_1 \}$, not $\{a_0, b_0\}$. That is, there are many choices of flux values that yield a small $\Lambda_{\rm ph}$, but the particular choice $\{a_0, b_0\}$ giving a small $\Lambda_0$ is not one of them. This means, with respect to $P(\Lambda_{\rm ph})$, the choice $\{a_0, b_0\}$ is not statistically preferred. As long as $P(\Lambda_{\rm ph})$ continues to peak at $\Lambda_{\rm ph}=0$, preference for small $\Lambda_{\rm ph}$ will continue to hold, irrespective how many loops we include.
In this sense, the statistical preference approach simply bypasses the radiative instability problem. 

This way of bypassing the radiative instability problem should also apply to higher order radiative corrections. It should also apply to the masses as well when the probability distribution $P(m^2)$ for some scalar mass also peaks at $m^2=0$. Furthermore, one may convince oneself that this statistical preference for a small $\Lambda$ also bypasses the disruptions caused by phase transitions during the evolution of the early universe, as the universe rolls down the landscape in search of a meta-stable minimum.

Actually we are interested only in the preferred value of the physical $\Lambda$. However, including quantum effects fully is in general a very challenging problem in any theory. Fortunately, if one can argue that the peaking behavior of $P(\Lambda)$ is hardly modified by quantum corrections, as this model suggests, a simpler tree-level result provides valuable information on the statistical preference of a small physical $\Lambda$. For ground states in string theory, an effective potential description may be sufficient to capture the physics of the value of $\Lambda$ in some region of the landscape.  We may hope that stringy corrections will not qualitatively disrupt the statistical preference approach adopted here.

\subsection{Finite Temperature $T$ and Phase Transition}
 
 Suppose the Universe starts out at a random point somewhere high up in the landscape, at zero temperature (for zero temperature, we mean zero thermal temperature, not the Gibbons-Hawking temperature ${H}/{2\pi}={\sqrt{V}}/{2\pi\sqrt{3} M_p}$, which is assumed to be negligible here). It rolls down and ends up in a local minimum. Because it starts from a random point, this minimum may be considered to be randomly chosen. If most of the vacua have a small $\Lambda$, it is likely that this minimum is one of these small $\Lambda$ vacua.
 
 
 What happens if we turn on a finite temperature $T$ ? We have essentially the same landscape (see below), but is starting from a different point up in the landscape, so the evolution of the Universe will be different and possibly ending at a different local minimum, also randomly chosen. As temperature $T \rightarrow 0$, we find that the chosen local vacuum at $T$ probably turns out to have a small $\Lambda$ at $T=0$, because most vacua at $T=0$ have a small $\Lambda$. 
If the chosen local vacuum has a critical temperature $T_c < T$, phase transition happens as $T$ drops below $T_c$. If this is a second order phase transition, then the Universe will roll away to another local minimum, which is likely to have a small $\Lambda$ as $T\rightarrow 0$, because most vacua at $T=0$ have a small $\Lambda$. If it is a first order phase transition, the Universe will stay at this vacuum as $T\rightarrow 0$ (before tunneling). This vacuum should have a small $\Lambda$, because most vacua at $T=0$ have a small $\Lambda$. In all cases, we see that the Universe most likely end up in a vacuum with a small  $\Lambda$. It is possible that this same vacuum has a relatively large $\Lambda$ at finite $T$. As an illustration, let us go back to the $\phi^4$ model and its mini-landscape. \\
 
 Since we sweep through the ``flux" parameters in $V(\phi)$ (\ref{A11}, \ref{phi4d}), we have in effect included cases both before and after spontaneous breaking. 
  Let us consider two possibilities here. 
  
 (1)  Suppose at finite temperature $T$, we have
 \begin{equation}
  \label{A1T}
V_0(\phi, T) = a \phi +\frac{(gT^2-b)}{2} \phi^2 + {c \over 3!} \phi^3 +\frac{d}{4!}\phi^4
 \end{equation}
 where $g$ is a calculable constant. Here, a $\phi$-independent but temperature-dependent term is ignored, since cosmologically it is not part of the dark energy; it contributes to the radiation energy density, which decreases as the universe expands and vanishes as $T \rightarrow 0$.
 
Let $b'=(b-gT^2) \in [-1, +1]$, so we can have $b>0$ while $b'<0$. That is, the mini-landscape already covers both the ``before and after spontaneous breaking" cases. (This is clearer if we look at the point where $a=c=0$.) We may choose to treat the finite temperature case as the landscape with the ranges of parameters slightly shifted. (Here, $b \in [-1, +1]$ $\rightarrow$ $b' \in [-1-gT^2, +1- gT^2]$.) Since the peaking (the divergence) of $P(\Lambda)$ at $\Lambda=0$ is unchanged if we shift a little the range of $b$, we see that  the preference for small $\Lambda$ is present both with or without the finite temperature effect.  Note that enlarging the range of $b$ does not impact on the peaking of $P(\Lambda)$ at $\Lambda=0$. Such a change will only change a little $P(\Lambda)$ away from $\Lambda=0$.

(2) For any of the parameters in $V(\phi)$ (\ref{phi4d}) to mimic a flux parameter, its magnitude should be fixed. To be specific, let us consider $b = qn$ where the magnitude $q$ is the ``charge", with dimension mass squared of order $M_s^2$, and integer $n=0, \pm 1, \pm 2, . . . . $. For a dense discretuum, we have taken $b\in[-1, 1]$, which includes a relatively large range of $n$ if $q$ is small enough. In string theory, $q$ is determined by some dynamics such as wrapping a cycle in the internal dimensions.  Implicitly, we have assumed that $q=U_{min}(\varphi)$, where
$U(\varphi)$ is the effective potential (at $T=0$) of another heavy modulus that has been integrated out. At finite temperature $T$, $q'=U_{min}(\varphi, T)\ne U_{min}(\varphi, 0)$. Here, $b'=q'n \ne qn$. This effectively changes the range of $n$ if we maintain $b\in[-1, 1]$. However, this has no impact on the peaking (the divergence) of $P(\Lambda)$ at $\Lambda=0$.

Overall, the finite temperature effect and possible phase transition are already built in the landscape picture. Sweeping through different temperatures is equivalent to sweeping the ``flux" parameters over some ranges. The peaking (the divergence) of the probability distribution $P(\Lambda)$ at $\Lambda=0$ is robust under these types of finite temperature effects, although $P(\Lambda)$ away from $\Lambda=0$ may be modified if we have to extend or shift the parameters' ranges.

\section{A K\"ahler Uplift Model of Flux Compactification}

Here we review a flux compactification model where the AdS vacua are K\"ahler uplifted to dS vacua
via the presence of an $\alpha'^3$ correction plus a non-perturbative term \cite{Rummel:2011cd}. Using reasonable probability distributions for the flux values, it has been shown in Ref\cite{Sumitomo:2012vx} that the probability distribution $P(\Lambda)$ peaks sharply at $\Lambda=0$, resulting in a median $\Lambda$ comparable to the observed value if the number of complex structure moduli $h^{2,1} \sim {\cal O}(100)$. We also summarize here the formulae needed to determine the bosonic masses of the resulting vacua.

\subsection{A Flux Compactification Model in Type IIB String Theory}

To be specific, consider a Calabi-Yau-like three-fold $M$ with a single ($h^{1,1}=1$) K\"ahler modulus and a relatively large $h^{2,1}$ number of complex structure moduli, so the manifold $M$ has Euler number $\chi(M)=2(h^{1,1}-h^{2,1}) <0$.
The simplified model of interest is motivated by orientifolded orbifolds \cite{Lust:2005dy,Lust:2006zg}, given by, setting $M_P=1$,
\begin{equation}
 \begin{split}
  V =& e^{K} \left(K^{I \bar{J}} D_I W D_{\bar{J}} {\overline W} - 3\left|W \right|^2\right),\\
  K =& K_{\rm K} + K_{\rm d} + K_{\rm cs}= -2    \ln \left({\cal V} + {\hat{\xi} \over 2} \right) -    \ln \left(S+\bar{S} \right) -  \sum_{i=1}^{h^{2,1}} \ln \left(U_i + \bar{U}_i  \right),\\
  {\cal V} \equiv& {\vol \over \alpha'^3 } =  (T + \bar{T})^{3/2},  \quad 
  \hat{\xi} =  -\frac{\zeta(3)}{4\sqrt{2}(2\pi)^3} \chi(M) \left( S + \bar{S} \right)^{3/2}>0, \\
  W =&  W_0(U_i,S) +  A e^{-a T}, \\
   W_0(U_i,S) =&  c_1 +\sum_{i=1}^{h^{2,1}} b_i U_i - S \left(c_2 + \sum_{i=1}^{h^{2,1}} d_i U_i\right) 
   +\sum_{i,j}^{h^{2,1}} \alpha_{ij}U_iU_j,
 \end{split}
 \label{LVS}
\end{equation}
The flux contribution to $W_0 (U_i,S)$ depends on the dilation $S$ and the $h^{2,1}$ complex structure moduli $U_i$ ($i=1,2,..., h^{2,1}$), while the non-perturbative term for the K\"ahler modulus $T$ is introduced in the superpotential $W$ \cite{Kachru:2003aw}. The dependence of $A$ on $U_i, S$ are suppressed.
The model also includes the $\alpha'$-correction (the $\hat{\xi}$ term) to the K\"ahler potential \cite{Becker:2002nn,Bonetti:2016dqh}, where $c_i, b_i$,  $d_i$ and $\alpha_{ij}=\alpha_{ji}$ are (real) flux parameters that may be treated as independent random variables with smooth probability distributions that allow the zero values.  

Note that the K\"ahler potential in terms of complex structure moduli for certain manifolds with $h^{2,1} =3$ is known, but its extension  for $h^{2,1} >3$ takes a form too complicated for us to see the interesting underlying properties. The simple extension adopted below allows us to solve this model semi-analytically to find the behavior of $P(\Lambda)$. In this sense, the model is at best semi-realistic. This form of the K\"ahler potential leads to $e^K \sim (\prod \mathrm{Re} U_i)^{-1}$ in the potential which is responsible to produce a small width in the peaking of $P(\Lambda)$.
Here we are interested in the physical $\Lambda$ (instead of, say, the bare $\Lambda$), so the model should include all appropriate non-perturbative effects, $\alpha'$ corrections as well as radiative corrections.
We see that the above simplified model (\ref{LVS}) includes a non-perturbative $A$ term to stabilize the K\"ahler modulus and the $\alpha'$ correction $\hat \xi$ term to lift the solution to de-Sitter space. In the same spirit, all  parameters in the model, in particular the coupling parameters $c_i, b_i$, $d_i$ and $\alpha_{ij}$ in $W_0$ (\ref{LVS}), should be treated as physical parameters that have included all relevant corrections.
Similar models have been proposed for the Large Volume Scenario \cite{Balasubramanian:2005zx} (see also \cite{Conlon:2005ki,Cicoli:2008va,Gray:2012jy}), and has been further analyzed in the search of de-Sitter vacua \cite{Balasubramanian:2004uy,Westphal:2006tn,Rummel:2011cd,deAlwis:2011dp}. Some explanations and justifications of the simplifications and approximations made can be found in Ref\cite{Rummel:2011cd,Sumitomo:2012vx}.

Before introducing the $A$ term for  K\"ahler modulus stabilization and the $\alpha'$ correction ${\hat \xi}$ term for K\"ahler uplift, supersymmetric solutions are obtained with $D_JW_0=\partial_JW_0 + (\partial_JK)W_0=0$ for each $J$ where 
\begin{align}
D_S W_0 &= -c_2-\sum d_iU_i - \frac{1}{S+\bar{S}}W_0, \nonumber \\
D_i W_0 & = b_i -S d_i  + 2\sum_j \alpha_{ij}U_j - \frac{1}{U_i + \bar{U}_i}W_0, 
\label{DWsu}
\end{align}
where $i=1,2,..., n=h^{2,1}$.
Let $S=s+i \nu_0$ and $U_j = u_j +i \nu_j$. For fixed flux values $b_j, c_j$, $d_j$ and $\alpha_{ij}$, which we take real values to simplify the analysis, we first solve for $D_JW_0=0$ to determine $u_i, s$ in terms of the flux values to yield $W_0= \omega_0 (b_j, c_j, d_j, \alpha_{ij}, s, u_i)= \omega_0 (b_j, c_j, d_j, \alpha_{ij})$ and insert this into $V$ (\ref{LVS}) to solve for $T$.

To simplify, let all real flux values be fixed, so $D_JW_0=0$ immediately give,
 \begin{align}
 \label{mc4a}
    v & \equiv  vf_1+2 r_1u_i=  vf_1+2 r_2u_2=  \cdots   = vf_n +2r_nu_n , \nonumber \\
    &f_i=(b_i-sd_i)u_i/v, \quad r_i= \sum_j \alpha_{ij} u_j , \nonumber \\
    & \nu_j=0,
  \end{align}
and the $u_i$ are solved in terms of $s$ and one of them, say $u_1$, or equivalently, $v$. Going back to Eq(\ref{DWsu}) allows us to solve for $v$ and $s$ in terms of the fluxes, and
\begin{align}
\omega_0 &= W_0|_{sol}= -2\left(sc_2 +\sum_i v(p_i-f_i)\right) = 2v , \nonumber \\
& p_i=(b_i+sd_i)u_i/v, \quad p=\sum p_i.
\end{align} 

Next we insert $\omega_0$ into the system and solve for $T$ that minimizes $V$ at its stable value in the presence of the $\alpha'$ correction ${\hat \xi}$ term. Since the imaginary part of the K\"ahler  modulus $T$ has a cosine type of potential, the extremal condition for this direction is satisfied when ${\rm Im} \, T = 0$.
Therefore we focus only on real part $t \equiv {\rm Re}\, T$. 
Since the $e^{-2a t}$ term is more suppressed than the $e^{-a t}$ term, we shall ignore it to obtain  \cite{Rummel:2011cd}
\begin{equation}
 \begin{split}
  {V  } &\simeq  e^K(-4\omega_0 A x^3) \left({2C \over 9 x^{9/2}} - {e^{-x} \over x^2} \right)= 4e^K (-\omega_0 Ax^3)  Y(x), \\
  e^K &= \frac{1}{(2t)^3 2s \prod 2u_i}, \quad C =  \frac{-27 \omega_0 \hat{\xi} a^{3/2}}{64 \sqrt{2}  A}, \quad
  x = a t.
 \end{split}
 \label{eq:westphal potential}
\end{equation}
The stability condition $\partial_{x}^2 V >0$ at the extrema $\partial_{x} V =0$ with respect to $x$ is easy to analyze, and we get the parameter range for stable positive $\Lambda$:
\begin{equation}
C_0 \lesssim  C < C_1 \quad  \to \quad   3.65 \lesssim C < 3.89,
\label{Cbound}
\end{equation}
where the lower bound is given by positivity of the minimum of $V$, while the upper bound is given by the stability constraint. Although we do not know the functional form for $A(S, U_i)$, $A$ depends on the flux values after $S$ and $U_i$ have been solved in terms of the flux parameters. So we shall simply treat $A$ as a variable that takes a range of values, including values so that $C$ satisfies the constraint (\ref{Cbound}), which in turn results in a bound on $Y(x)$,
\begin{equation}
0  \le  Y(x) < 6 \times 10^{-4}.
  \label{stable region for single Kahler}
\end{equation}

\begin{figure}[h] \label{fig:V_H}
 \begin{center}
  \includegraphics[scale=.7]{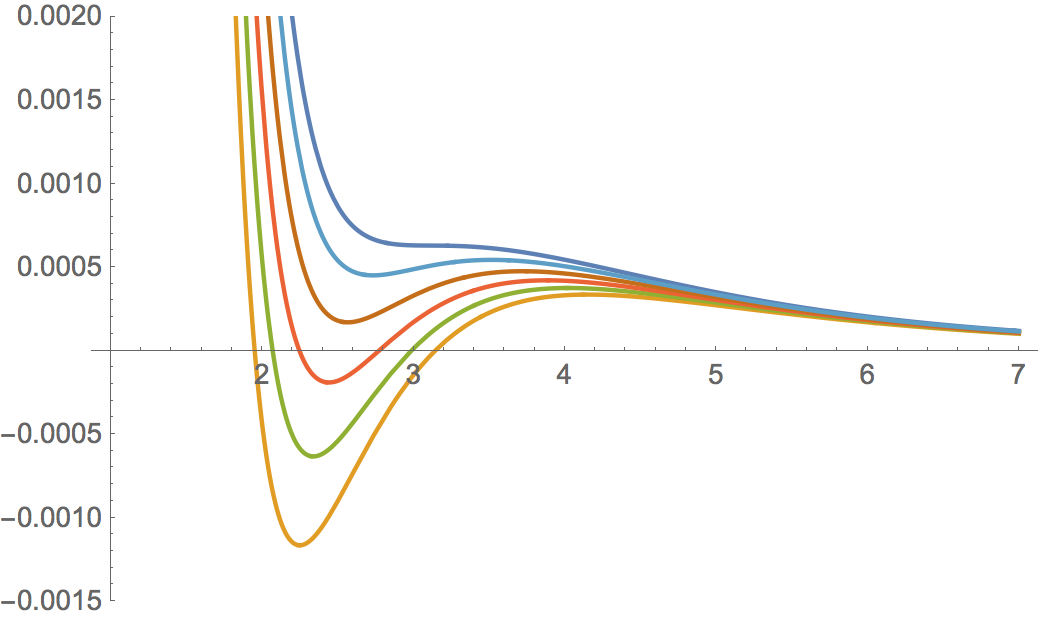}
 \end{center}
 \caption{The K\"ahler uplift of the potential $V(x)$ for different values of $C$.}
\end{figure}

If we satisfy the combination of parameters $C$ inside this region, with appropriate choice of $A$ and flux values, there is a stable solution in the range $2.50 \lesssim x < 3.11$ at $\Lambda \ge 0$. Up to an overall factor, the potential $V$ (\ref{eq:westphal potential}) is shown in Figure \ref{fig:V_H}. Solving $x$ in the allowed range for the minimum of $V$ (\ref{eq:westphal potential}), we finally obtain

\begin{align}
\Lambda &\simeq 4 e^K (-\omega_0 Ax^3) Y(x) , \quad  2.50 \lesssim x < 3.11.
\label{Lambda1}
\end{align}

\subsection{Moduli Masses}

Before adding non-perturbative terms and uplifting, the model has a no-scale structure, so
\begin{align}
K^{T\bar{T}}K_TK_{\bar{T}}=3, \quad V=e^K(K^{S\bar{S}}D_S W D_{\bar{S}}\overline{W}+K^{U_i\bar{U}_j}D_{U_i} W D_{\bar{U}_j}\overline{W} ).
\end{align}
Since $V$ has the form of a perfect square, the $S$ and $U_i$ masses are semi-positive. Let us find their masses now.

Since the kinetic terms are $K_{I\bar{J}}\partial_{\mu}\Phi^I\partial^{\mu}\Phi^{\bar{J}}$ (which are diagonal in $S,U_i$), we have the following canonically normalized mass-square matrix,
\begin{align}
m^2_{ss} \big|_{\rm min}= \frac{1}{2K_{S\bar{S}}}\partial_s^2 V, \quad m^2_{su_i}\big|_{\rm min} = \frac{1}{2\sqrt{K_{S\bar{S}}K_{U_i\bar{U}_i}}} \partial_s \partial_{u_i}V, \quad  m^2_{u_i u_j}\big|_{\rm min} = \frac{1}{2\sqrt{K_{U_i\bar{U}_i}K_{U_j\bar{U}_j}}} \partial_{u_i} \partial_{u_j}V,
\end{align}

The terms $D_iW_0$ in the potential are vanishing at minimum unless all $D_iW_0$ are hit by the derivative. The resulting $(n+1) \times (n+1)$ mass-square matrix $m^2_{ij}$ is simply (with  $s=u_0$, $i=0, 1, 2, \cdot \cdot \cdot, n$)
\begin{align}
\label{mass2f}
 m^2_{ij} &= e^K 4 u_i u_j\left[ \partial_i G_0 \partial_j G_0 + \sum_{k} \partial_i G_k \partial_j G_k  \right]  \nonumber \\
 &=4e^K H_{ij} = 4e^K \sum_k F_{ik} F_{jk}, \nonumber \\
 F_{ik} &= u_i \partial_i G_k, \quad  G_0 = -2sD_SW_0, \quad  G_j =  -2u_j D_{U_j} W_0,
\end{align}
where $G_k$ are given in terms of $D_JW_0$ (\ref{DWsu}). We see that the Hessian $H$ can be written as the product of the $(n+1) \times (n+1)$ matrix $F$ and its transpose $F^T$. Since we expect $n \sim {\cal O}(100)$, we like to present the analysis in two steps.

The case where $\alpha_{ij}=0$ has been studied in some detail, so let us consider this simplified case first. Before introducing the $A$ term for  K\"ahler stabilization and the $\alpha'$ correction ${\hat \xi}$ term for K\"ahler uplift, supersymmetric solutions are obtained with 
\begin{align}
D_S W_0 = -c_2-\sum d_iU_i - \frac{1}{S+\bar{S}}\bigg(c_1 - S c_2 + \sum(b_i-S d_i)U_i\bigg), \nonumber \\
D_i W_0 = b_i -S d_i  - \frac{1}{U_i + \bar{U}_i}\bigg(c_1 - S c_2 + \sum_{j}(b_j-S d_j)U_j \bigg).
\label{DWs}
\end{align}
Assuming all flux values to be real, so we obtain, for each $u_i$,
 \begin{equation}
 \label{mc4}
    v\equiv vf_i = (b_i -s d_i)u_i,
  \end{equation}
and the $u_i$ are solved in terms of $s$ and one of them, say $u_1$, or equivalently, $v$. Going back to Eq(\ref{DWs}) allows us to solve for $v$ and $s$ in terms of the fluxes.
So we have $W_0$ solved, for $n> 2$, 
    \begin{equation}
   \label{mc6}
    \omega_0=W_0|_{\rm min} =  -2 {c_1 - s c_2\over n-2} =2v.
  \end{equation}

\subsection{Preference for an Exponentially Small Cosmological Constant}

Now we sweep through the flux values $c_j, b_j$ and $d_j$ treating them as independent random variables (or a variation way of sweeping) to find the probability distribution $P(\Lambda)$. The ranges of flux values are constrained by our weak coupling approximation (i.e., $s>1$) et. al.. For any reasonable probability distributions $P_i(c_j)$, $P_i(b_j)$ and $P_i(d_j)$, we find that $P(\Lambda)$ peaks (and diverges) at $\Lambda=0$. To quantify this peaking behavior,  it is convenient to summarize the result by looking at $\Lambda_{Y \%}$.  That is, there is $Y \%$ probability that $\Lambda_{Y \%} \ge \Lambda  \ge 0$. So $\Lambda_{50 \%}$ is simply the median. 
There we find that,  as a function of the number $h^{2,1}$ of complex structure moduli, for $h^{2,1} > 5$ and $\Lambda \ge 0$,
\begin{align}
\label{lambdam}
\Lambda_{50 \%} &\simeq 10^{-h^{2,1} -2} M_P^4, \nonumber  \\
\Lambda_{10 \%} &\simeq 10^{-1.3 h^{2,1} -3} M_P^4, \nonumber\\
\langle\Lambda \rangle& \simeq 10^{-0.03 h^{2,1} -6} M_P^4,
\end{align}
where we have also given $\Lambda_{10 \%}$. We see that the average $\langle\Lambda \rangle$ does not drop much, since a few relatively large $\Lambda$s dominate the average value.
A typical flux compactification can have dozens or even hundreds of $h^{2,1}$, so we see that a $\Lambda$ as small as that observed in nature can be dynamically preferred. 
Note that the median of $\omega_0$ decreases very slowly as $h^{2,1}$ increases, for $h^{2,1} >10$ \cite{Sumitomo:2012vx},
\begin{equation}
\omega_{0, 50\%} \sim 10^{-2.1 - 0.03 h^{2,1}}.
\label{omegaB}
\end{equation}

For a vacuum taking the observed value of $\Lambda$ (\ref{Lambda1}) without fine-tuning $Y(x)$, we see that $e^K$ must be exponentially small, since $\omega_0 \sim 10^{-5}$ for $h^{2,1} \sim 100$ and $A$ has comparable order of magnitude value as $\omega_0$ because of the bound on $C$ (\ref{Cbound}).  
Comparing the $e^K$ factor (\ref{eq:westphal potential}) to $\Lambda$ (\ref{Lambda1}), we see that for the observed $\Lambda$ (\ref{L1}) with $h^{2,1} \sim 100$, a typical $u_i \sim 5$.

\section{Complex Structure Moduli Masses}

Having real flux variables $c_1,c_2,b_i,d_i$ with vanishing imaginary part, the potential without non-perturbative terms is given by
\begin{align}
V= \frac{1}{(2t)^32^{h_{2,1}+1}s \prod u_i}\bigg[ \big|2s(c_2+\sum_id_iU_i)+ c_1- sc_2 + \sum_i(b_i-Sd_i)U_i \big|^2 \nonumber\\
+ \sum_k\big| c_1-sc_2 - 2(b_k-sd_k)u_k + \sum_{j}(b_j-Sd_j)U_j \big|^2 \bigg] ,\\
S=s+i\sigma, \quad U_j = u_j +i \nu_j  . \nonumber
\end{align}

First, one notices that the Hessian (mass matrix) for the real parts does not mix with the Hessian for the pseudo-scalar parts, so we can analyze them separately. 
Next we see that the mass matrix for $s$ and $u_i$ can be rewritten in the following more compact form
\begin{align} \label{mreal}
m^2_{ss} &= 4e^Kv^2\bigg( 1 +\sum_k p_k^2 \bigg)=4e^Kv^2\bigg( 1 + q \bigg), \nonumber \\
m^2_{s i} &=4e^K v^2 \bigg(  \sum_k p_k -3p_i \bigg)=4e^K v^2 \bigg( p -3p_i \bigg), \nonumber \\
m^2_{ij} &=4e^Kv^2\bigg( 4 \delta_{ij} + (n-4)+p_i p_j \bigg), \nonumber \\
&q =\sum_k p_k^2, \quad p = \sum_k p_k. 
\end{align}
where $v={w_0}/{2}=(b_i -sd_i)u_i$ and $p_i = (b_i +sd_i)u_i/v$. 
So the characteristic equation of the matrix $e^{-K} m^2/(4v^2)$ is, with $n=h^{2,1}$ (see Appendix \ref{AppA}),
\begin{align}
(4-\lambda)^{n-2} & \left[( q -\lambda)  \begin{vmatrix}
q +4-\lambda & (n-4)p \\  p &n(n-4) +4-\lambda \end{vmatrix} \right.   \nonumber \\
& \quad \quad \quad \left.+ \begin{vmatrix}
3p^2 -8q +4-\lambda & (3n-8)p^2 \\ 3q -p^2 +n-4 & n(n-4) -(n-3)p^2 +4-\lambda 
\end{vmatrix} \right],   
\label{Charaeq}
\end{align} 
for the eigenvalue $\lambda$.
Let us label the mass eigenmodes as $\varphi_i$ ($i=1,2,..., (n+1$)), so the first two are heaviest, with masses as $m_{s1} \ge m_{s2}$, while $\varphi_3$ has mass $m_{s3}$ and the remaining  $\varphi_i$ have the same degenerate mass
\begin{equation}
 m_{s4}= m_{s5}=.... =m_{s(n+1)}= e^K16v^2 =4e^K\omega_0^2. 
\end{equation}
That is, the ($n -2$) number of real moduli have twice the gravitino mass.
  
Let us first take an order-of-magnitude look at this degenerate mass. 
Comparing with $\Lambda$ (\ref{Lambda1}), we have
\begin{equation}
\frac{m_{s4}^2}{\Lambda} =\frac{m_{s4}^2M_P^2}{\Lambda} =  \frac{- \omega_0 }{Ax^3Y(x)} = \frac{64 \sqrt{2} C}{27 {\hat \xi}a^{3/2}x^3 Y(x)} .
\end{equation}
Since $C$, $x$, $a$ and $\hat \xi$ are either bounded or have typical order-one values, while $Y(x)$ (\ref{stable region for single Kahler}) is also bounded, we see that 
$$m^2M_P^2 /\Lambda \sim {\cal O}(1),$$
when $\Lambda$ takes the observed value, although one may fine-tune $Y(x) \sim 10^{-20}$ (note the allowed range of $Y(x)$ (\ref{stable region for single Kahler})) if we want these moduli to play the role of light dark matter. Alternatively, we can turn on the quadratic couplings $\alpha_{ij}$ among the complex structure moduli in $W_0$ (\ref{LVS}) to raise their masses, to which we shall discuss in the next section.

Let us now look at the masses of the remaining 3 heavier scalars. We are left with a characteristic equation (\ref{Charaeq}) which is cubic in the eigenvalue $\lambda$, so it can be analytically solved. For $n=h^{2,1} \sim 100$, $|p| \ll h^{2,1} \lesssim \sqrt{q}$, so they have approximate masses given by 
 \begin{align}
 m_{s1}^2 \simeq & \, (q +4\sqrt{q}) e^K\omega_0^2 ,\nonumber \\
 m_{s2}^2 \simeq & \, (q - 4\sqrt{q}) e^K\omega_0^2 ,\nonumber \\
 m_{s3}^2 \simeq & \, (h^{2,1}-2)^2  e^K\omega_0^2. 
 \end{align}
Here the mass of $\varphi_3$ can be heavier than $\varphi_4$ by up to about 2 orders of magnitude.
The masses of the heaviest 2 moduli increase as $n$ increases. Numerically, we see that they can take a range of values, with the mean values ($m_{s1} \sim m_{s2}$) going like
\begin{equation}
r_m=m_{s1}^2M_P^2/\Lambda \sim 10^{(0.12 \pm 0.05)n + 3 \pm 2} ,
\end{equation}
where the coefficient of $n=h^{2,1}$ is obtained numerically, with $0.12$ for the median, $r_m^{50\%}$, $0.12+0.05$ for $r_m^{75\%}$ and $0.12-0.05$ for $r_m^{25\%}$. The other exponent factor $3\pm 2$ comes from estimates of the remaining factors without fine-tuning. Comparing this to $\Lambda$ (\ref{lambdam}), where $n \sim 120$ is reasonable, we see that $m_{s1} \sim m_{s2}$ can have masses in the range for dark matter. Their self-couplings are also very small.
However, these two heavy moduli contain significant components of the dilaton (while the others have negligible contributions from the dilaton). To avoid modifying the gravitational force via dilaton exchange, we may like them to have mass values higher than appropriate as dark matter candidates.
  
As explained in Ref\cite{Rummel:2011cd,Sumitomo:2012vx}, K\"ahler uplift will have little impact on these moduli masses.
Going back to $V$ (\ref{eq:westphal potential}), we see that the overall factor $e^K$ means all masses and couplings will be exponentially suppressed, much like the suppression of $\Lambda$. Within this simple framework, any Higgs field introduced will probably have masses much like the moduli masses, which is much too small for the observed Higgs boson in the electroweak theory.
Clearly we have to consider string theory scenarios with more structure to have multiple mass scales to fit nature. We shall come back to this point later.

\subsection{The Axion Masses}

Let us now look at the axion masses $m_{ai}$. The mass matrix for axions can be obtained in a similar way as given in Appendix \ref{AppA}. Recall that $S=s+i\sigma$ and $U_j = u_j +i \nu_j$, 
 the axion mass matrix is given by
\begin{align}
m^2_{\sigma} &= 4e^K v^2\bigg( 1 +\sum_k p_k^2 \bigg), \nonumber \\
m^2_{\sigma \nu_i} &=4e^K v^2 \bigg(  \sum_k p_k +p_i \bigg), \nonumber \\
m^2_{\nu_i\nu_j} &=4e^Kv^2 \bigg( n+p_i p_j \bigg),
\end{align}
With the above way of finding characteristic equation, one can immediately see that there are $(n-2)$ massless axions. The characteristic equation of $e^{-K} m_a^2/(4v^2)$ is,
\begin{equation}
(-\lambda)^{n-2} \left[( q -\lambda)\begin{vmatrix}
q -\lambda & np \\ p& n^2 -\lambda\end{vmatrix} + \begin{vmatrix}
-p^2 -\lambda & -n p^2 \\ -p^2 -q +n & -(n+1)p^2 -\lambda +n^2\end{vmatrix} \right].
\label{Charaeqa}
\end{equation}

We find that the axion masses are
 \begin{align}
 m_{a1}^2 \simeq & \, q e^K\omega_0^2, \nonumber \\
 m_{a2}^2 \simeq & \,q e^K\omega_0^2 , \nonumber \\
 m_{a3}^2 \simeq &\, n^2  e^K\omega_0^2, \nonumber \\
 m_{a4}^2 =& m_{a5}^2 = ... =m_{an}^2 = 0.
 \end{align}
The masses of the 3 massive bosons have values comparable to the 3 corresponding heavy scalars.
In the $(n-2)$ massless directions, there are positive quartic terms so that the vacuum is stabilized.
In general, we expect the axion masses to be uplifted via non-perturbative terms, of the form $A_k(S,U_j)e^{-a_kU_k}$, which can be introduced into the superpotential $W_0$ (\ref{LVS}). 
In general, we expect an instanton effect generates a term of the form
\begin{equation}
V(a) = m_a^2f_a^2 \left(1 -\cos(a/f_a) \right)
\end{equation}
where $f_a$ is the axion decay constant or coupling parameter of the axion $a$.

 \subsection{Lifting the Complex Structure Moduli Masses}

Let us now turn on the $\alpha_{ij}$ couplings in $W_0$ (\ref{LVS}) step by step. First, we note that $m_{s1} \sim m_{s2} \gg m_{s4}$ because of the dilaton $S$ couplings $d_i$ to the $U_i$ in $W_0$. If we have instead set the flux parameters $d_i=0$ while keeping $n-1$ number of couplings $\alpha_{1j}$ ($j=2, 3, \cdot \cdot \cdot, n$) as the non-zero flux parameters, then the roles of $S$ and $U_1$ interchange and  the relatively heavy bosons would be the two complex structure moduli that contain most of $u_1$.  

Let us turn on $\alpha_{ij}$ step by step.

First turn on only one coupling $\alpha_{11}$; here we see that the $(n+1) \times (n+1)$  matrix  $F$ (\ref{mass2f})  goes from 
 \begin{align}
F_0=F(\alpha_{ij}=0) =\begin{pmatrix} -1 & p_1 & p_2 & p_3 & p_4 & \cdot \cdot \cdot \\
p_1 & -1 & +1 & +1 &+1 & \cdot \cdot \cdot \\
p_2 & +1 & -1 & +1 &+1 & \cdot \cdot \cdot \\
p_3 & +1 & +1 & -1 &+1 & \cdot \cdot \cdot \\
\cdot \cdot \cdot &&&&& \\
\cdot \cdot  &&&&&
\end{pmatrix}
\end{align}
to
 \begin{align}
F_1=F(\alpha_{11})=\begin{pmatrix} -1 & x_1 & p_2 & p_3 & p_4 & \cdot \cdot \cdot \\
x_1 & f_1-2-y_1 & f_1 & f_1 & f_1 & \cdot \cdot \cdot \\
p_2 & +1 & -1 & +1 &+1 & \cdot \cdot \cdot \\
p_3 & +1 & +1 & -1 &+1 & \cdot \cdot \cdot \\
\cdot \cdot \cdot &&&&& \\
\cdot \cdot  &&&&&
\end{pmatrix}
\end{align}
to
 \begin{align}
F_n=F(\alpha_{1j})=\begin{pmatrix} -1 & x_1 & p_2 & p_3 & p_4 & \cdot \cdot \cdot \\
x_1 & f_1-2-y_1 & f_1-y_2 & f_1-y_3 & f_1-y_4 & \cdot \cdot \cdot \\
p_2 & 1-y_2 & -1 & +1 &+1 & \cdot \cdot \cdot \\
p_3 & 1-y_3 & +1 & -1 &+1 & \cdot \cdot \cdot \\
\cdot \cdot \cdot &&&&& \\
\cdot \cdot  &&&&&
\end{pmatrix}
\end{align}
where we have turned on the $n$ couplings $\alpha_{1j}$ in the last $F=F_n$. Recall that $f_j=(b_j-sd_j)u_j/v$, $p_j=(b_j+d_j)u_j/v$, $y_j=4 \alpha_{1j}u_ju_1/v$ and $x_1=p_1 +2\alpha_{1j}u_ju_1/v=p_1+y_1/2$. Note that  $f_1+2\alpha_{1j}u_ju_1/v=1$.

Recall that $F_0$ yields $(n-2)$ degenerate masses with 3 heavier bosons. It is easy to see that $F_1$ will yield $(n-3)$ degenerate masses with 4 heavier bosons while $F_n$ will yield $(n-4)$ degenerate masses with 5 heavier bosons. Turning on more $\alpha_{ij}$ couplings will lift more of the degenerate masses to heavier values. Numerically, we see that having masses of order (\ref{dark1}) suitable for dark matter without fine-tuning is quite easy.

\subsubsection{Example}
 
Consider adding a term $U_1\sum_i\alpha_{i}U_i$ in the superpotential,
\begin{align}
W_0 &= c_1-Sc_2 +\sum_i(b_i-Sd_i)U_i + U_1\sum_{i}\alpha_{i}U_i, \quad V=e^K(H\bar{H} + \sum_k G_k \bar{G}_k), \nonumber \\
H &=\sqrt{K^{S\bar{S}}}D_S W_0 = -2s(c_2 +\sum_i d_iU_i)-W_0, \nonumber \\
G_k  &= \sqrt{K^{U_k\bar{U}_k}}D_{U_k} W_0 = 2u_k(b_k -Sd_k +U_1\alpha_k + \delta_{1k}\sum_j\alpha_jU_j ) -W_0.
\end{align}
\begin{align}
 m_{ss}^2& = 4 e^K\big[ v^2 + \sum_k(v+2sd_ku_k)^2\big], \nonumber \\
 m_{si}^2 &= 4 e^K \Big[ v(-v-2sd_iu_i )  +\sum_k (-v -2sd_ku_k)(2v\delta_{ki} +2\delta_{1i}\alpha_ku_iu_k +2\delta_{1k}\alpha_iu_iu_k-v) \Big] ,\nonumber \\
  m^2_{u_iu_j} &=4e^K\Big[(v+2sd_iu_i)(v+2sd_ju_j ) \nonumber\\
  & \quad +\sum_k(2v\delta_{ki} +2\delta_{1i}\alpha_ku_iu_k +2\delta_{1k}\alpha_iu_iu_k-v)(2v\delta_{kj} +2\delta_{1j}\alpha_ku_ju_k +2\delta_{1k}\alpha_ju_ju_k-v)   \Big].
\end{align} 
Using the same method of calculating the determinant, one can immediately see there are $(h_{2,1}-4)$ particles of the same mass $4 e^Kw_0^2$. Similar to the previous case, the axio-dialton (states with ${\cal O}(1)$ mixing with $S$) is the heaviest and it is separated from the scale of $\lambda$ by roughly $10^{0.12 h_{2,1}}$. Some particles become heavier compared to those for $\alpha_{ij}=0$ but the uplift from the scale of $\Lambda$ is not so big. We give an example for the case of $h_{2,1}=10$: the distribution of $\log_{10}(m^2M_P^2/\Lambda)$ for one of the uplifted mass in FIG 4. Note that there are also some much heavier boson mass samples in the tail.
\begin{figure}[h]
\label{FIG4}
 \begin{center}
  \includegraphics[scale=.7]{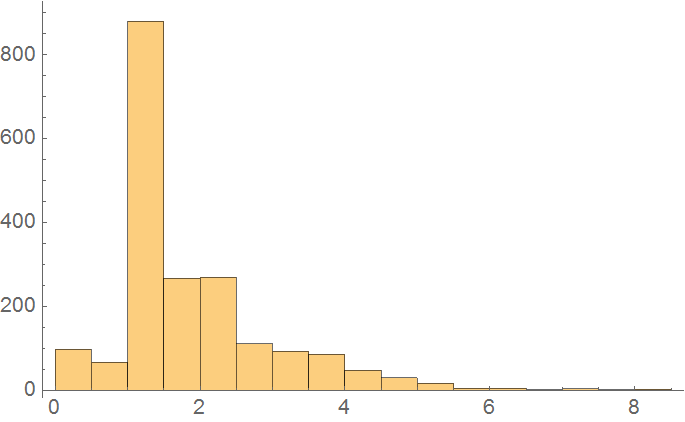}
 \end{center}
 \caption{Distribution of $\log_{10}(m^2M_P^2/\Lambda)$ for the uplifted mass due to the presence of non-zero $\alpha_i$. Here  $\alpha_i$ is randomized in $[-1,1]$.}
\end{figure}


\section{Moduli Masses in Racetrack K\"ahler Uplift}

The K\"ahler uplift model studied in the last section has a single non-perturbative term in the superpotential $W$. To relax the constraint on the volume size, we generalize the model to include two non-perturbative terms in $W$, i.e., the {\it racetrack} model. This model has been studied in {\cite{Westphal:2005yz,deAlwis:2011dp,Sumitomo:2013vla}}.
Unlike the K\"ahler uplift model studied previously, the $\alpha'$-correction is more controllable for the meta-stable de-Sitter vacua in the racetrack case since the constraint on the compactified volume size is very much relaxed.
So the model admits solutions with a large adjustable volume.

Interestingly, in this Racetrack K\"ahler uplift model, the stability condition for both the real and imaginary sectors requires that the minima of the potential $V$ always exist for $\Lambda \ge 0$  at large volumes.
Further, the cosmological constant $\Lambda$ is naturally exponentially suppressed as a function of the volume size, and the resultant probability distribution $P(\Lambda)$ for $\Lambda$ gets a sharply peaked behavior toward $\Lambda \rightarrow  0$, which can be highly diverging \cite{Sumitomo:2013vla}.
This peaked behavior of $P(\Lambda)$ can be much sharper than that of the previous K\"ahler Uplift model with a single non-perturbative term studied in \cite{Sumitomo:2012wa,Sumitomo:2012vx}. Getting an exponentially small median for $\Lambda$ is natural.

The racetrack K\"ahler uplift model is similar to the above K\"ahler Uplift model, but with one major addition. The super-potential  $W$ now has two non-perturbative terms for the K\"ahler modulus $T=t + i\tau$ instead of one,
 \begin{equation}
W= W_0 (U_i, S) + W_{\rm NP} = W_0 (U_i, S) + A e^{- a T} + B e^{-b T}, 
\end{equation}
where the coefficients $a=2\pi/N_1$ for $SU(N_1)$ gauge symmetry and $b=2\pi/N_2$ for $SU(N_2)$ gauge symmetry. 
In the large volume region and in units where $M_P=1$, the resulting potential may be approximated to
\begin{equation}
 \begin{split}
  &V \simeq \left(-{a^3 A \, W_0\,  \over 2}\right) \lambda (x,y), \\
  & \lambda (x,y) =  - {e^{-x} \over x^2} \cos y - {\beta \over z} {e^{-\beta x} \over x^2} \cos (\beta y) + {\hat{C} \over x^{9/2}},\\
  &x = a t, \quad y=a \tau, \quad   z = A/B, \quad \beta = b/a=N_1/N_2 >1, \quad {\hat{C}} = -{3 a^{3/2} W_0 \, \xi \over 32 \sqrt{2} A}.
 \end{split}
 \label{approximated potential}
\end{equation}
The extremal conditions $\partial_t V = \partial_{\tau} V = 0$ may be expressed as the relations:
\begin{equation}
 \begin{split}
  {1\over z} =& e^{(\beta-1) x} {- 2 x + 5 + 9 e^{x} x^2 \lambda \over \beta (2 \beta x - 5)},\quad
  {\hat{C}} =  2 e^{-x} x^{7/2} { (\beta-1) + e^{x} (x^2 + 2 x) \lambda \over 2 \beta x -5}.
 \end{split}
 \label{relations for extrema}
\end{equation}
The non vanishing Hessian (mass squared) components are,
\begin{align}
 \partial^2_{x} \lambda &= \frac{e^{-x} (\beta -1) \left(4 \beta  x^2-10 (\beta +1) x+35\right)-9 \lambda   x (\beta  x (2 \beta  x-3)-10)}{2 x^3 (2 \beta  x-5)}, \nonumber \\
  \partial^2_{y} \lambda & = \frac{e^{-x}(\beta-1) \left(-2 \beta  x+5 \left(\beta +1\right)\right)+9 \beta ^2 \lambda  x^2}{x^2 (2 \beta  x-5)}.
\end{align}
Requiring both of them to be positive (hence the extremum is a minimum) gives, 
\begin{equation}
 \label{exactcond}
  e^{-x} {(\beta-1) (2 \beta x -5 (\beta+1)) \over 9 \beta^2 x^2} \le \lambda \le e^{-x} {(\beta-1) (4 \beta x^2 - 10 (\beta+1) x + 35) \over 9 x (2 \beta^2 x^2 - 3 \beta x-10)}.
\end{equation}

The typical values of $a$, $\beta$ and $x$ are ${\cal O}(2\pi/16)$, ${\cal O}(1)$ and ${\cal O}(100)$ respectively, and the $e^{-x}$ factor suggests very small $\Lambda$ as well as moduli masses. After randomizing $W_0$, $A$ and $B$, we collect the solutions and find that
the probability distribution $P(\Lambda)$ for small positive $\Lambda$ is approximately given by \cite{Sumitomo:2013vla},
\begin{equation}
 P(\Lambda) \stackrel{\Lambda \rightarrow 0}{\sim} {243 \beta^{1/2} \over 16 (\beta-1)} {1 \over \Lambda^{\beta+1 \over 2 \beta}  (-\ln \Lambda)^{5/2}}.
  \label{asymptotics of PDF}
\end{equation}
So for $\beta \gtrsim 1$, we see that the diverging behavior of $P(\Lambda)$ is very peaked as  $\Lambda \rightarrow 0$.
Since $(\beta +1)/2 \beta < 1$, $P(\Lambda)$ is normalizable, i.e.,$\int P(\Lambda) d \Lambda =1$. It is informative to introduce the value $\Lambda_Y$ that $Y\%$ of the data fall within it : $\int_0^{\Lambda_Y} d\Lambda \, P(\Lambda) = Y\%$, where $\Lambda_{50}$ is the median. For illustration, we have 
\begin{align}
\beta &=1.10 : \quad \quad  \Lambda_{50} = 7.08 \times 10^{-10}, \quad \Lambda_{10} = 3.61 \times 10^{-24} , \\
 \beta &=1.04 :   \quad \quad  \Lambda_{50}= 5.47 \times 10^{-19}, \quad \Lambda_{10} = 2.83 \times 10^{-54}.
\end{align}
We also see that both $t$ and $\tau$ masses are exponentially suppressed. By using the above inequality (\ref{exactcond}) and the small value of $\Lambda$, we can obtain bounds on both masses,
\begin{align}
\frac{m^2_t}{\Lambda} &= \frac{\partial_t^2 V}{2K_{T\bar{T}}\Lambda} \leq  \frac{  9\beta x+30(\beta+1)}{a^4 (  2\beta x-5(\beta+1))}, \nonumber \\
\frac{m^2_{\tau}}{\Lambda} &= \frac{\partial_{\tau}^2 V}{2K_{T\bar{T}}\Lambda}\leq \frac{6 x ( 3 \beta x+10(\beta+1))}{a^4 \left(4 \beta  x^2-10 (\beta +1) x+35\right)}.
\end{align}
Solving for $x \sim {\cal O}(100)$, we see that the K\"ahler modulus masses are exponentially small unless one fine-tunes one of the denominating factor to a very small value.

As pointed out in Ref \cite{Goodman:2000tg,Guth:2014hsa,Fan:2016rda}, axions as light dark matter with weak repulsive self-coupling may possess interesting properties such as driving long range interactions while those with attractive self-interaction may lead to localized clumps. We demonstrate here axions with repulsive interaction can be constructed from the class of model considered here.  In general, an axion with a potential of the form :
$$V(a)= m^2f^2 \left(1-\cos(a/f)\right) \simeq \frac{m^2}{2}a^2 -\frac{m^2}{4!f^2}a^4 + ... $$
which yields an attractive self-coupling. Here, because (\ref{approximated potential}) has two cosine terms with opposite coefficients (in canonically normalized fields):
\begin{align}
V(a) &= V(0)+ \frac{m_1^2+m_2^2}{2}a^2 - \frac{1}{4!}\left(\frac{m_1^2}{f_1^2}+\frac{m_2^2}{f_2^2}\right)a^2+... \nonumber \\
  & = V(0)+\frac{-a^3 A  W_0}{ 4 K_{T\bar{T}}} \left(\frac{\beta ^3 e^{-\beta  x}}{2 x^2 z}+\frac{e^{-x}}{2 x^2}\right)a^2+\frac{-a^3 A  W_0}{ 8 K_{T\bar{T}}^2}\left(-\frac{\beta ^5 e^{-\beta x}}{24 x^2 z}-\frac{e^{-x}}{24 x^2}\right)a^4.
\end{align}
 We see that the resulting self-coupling can be repulsive and indeed it is in the parameter region of interest if it is a candidate for light dark matter.

\section{Discussions}

So far, we have a few looks at the global picture of some corners of the string landscape. As illustrated by the K\"ahler uplift models discussed, we see hints that, of the meta-stable solutions, most of them have very small $\Lambda$, while each such vacuum has very light bosons.  Here we like to discuss a few issues related to this property.

\subsection{Tunneling Suppression}

Let $M_P= G_N^{-1/2}$ and $M_{pl}=M_P/\sqrt{8 \pi}$.

Suppose it is a scalar boson
$$ V= \Lambda +  \frac{m^2}{2}\phi^2 + \frac{m^2}{4! f^2} \phi^4 +...$$
with barrier height $V_{bar} \sim \Lambda$.
With $\Lambda \sim 10^{-122} M_{P}^4$ and $m \sim 10^{-50}M_{P}$, so we have barrier wall tension 
$$\sigma \sim \Lambda/m \simeq 10^{-72} M_P^3.$$
Since there are vacua nearby that have comparable or smaller vacuum energy densities (say, one with $V_- \lesssim \Lambda$), tunneling via CdL is given by 
$$T \sim e^{-B}, \quad B_{CdL} \simeq \frac{27 \pi^2 \sigma^4}{2 \epsilon^3} \rightarrow \frac{2 \pi^2 \sigma}{H^3}$$
where $\epsilon \simeq \Lambda -V_-$ and the Hubble constant  
$$H^2 \simeq 8 \pi G_N \Lambda/3 = 8 \pi \Lambda/3M_{P}^2$$
while for Hawking-Moss,
$$B_{HM}=\frac{3M_P^4}{8} \left( \frac{1}{\Lambda} -\frac{1}{(\Lambda+V_{bar})}\right) 
\sim \frac{3M_P^4}{16}\frac{1}{\Lambda} \sim \frac{8\pi \Lambda}{3 H^4},$$
where $V_{bar} \sim \Lambda$.
We see that, in either case
$$B > 10^{110}$$ so tunneling out of such a low $\Lambda$ vacuum is very suppressed. \\

\subsection{Why Not AdS Vacuum ?}

In the Introduction, we envision the scenario how we might end up in a dS vacuum with a small $\Lambda$.
 Our universe rolling down the landscape after inflation is unlikely to be trapped by a relatively high dS vacuum, since there is hardly any around. So it rolls down towards the region with numerous low $\Lambda$ vacua. However, since it has to pass through the positive $\Lambda$ region first, it is likely to be trapped at a small positive $\Lambda$ vacuum before reaching any AdS vacua (as illustrated in FIG 1(a)). 

Once it reaches the low $\Lambda$ region, it tends to search for a minimum spot.
In an actual situation, it may roll in and then out of a $\Lambda$ vacuum if it has enough kinetic energy to move on \cite{Flanagan:1999dc}. This may happen a few times before finally, with the help of some damping, it ends up in the vacuum that our universe is sitting in today.  One may like to ask why we do not end up in an AdS vacuum. We do not have an answer to this possibility. However, it is interesting to note that tunneling to an AdS vacuum leads to a crunch, as shown in Ref\cite{Coleman:1980aw}. In this situation, we see that ${\dot \phi}$ blows up, showing that the tunneling to an AdS vacuum is unstable. 

Even if an AdS vacuum is stable against perturbing a modulus $\phi$, when its mass-squared $m^2 \ge 0$ (or not too negative), it is probably unstable against a non-linear perturbation involving its time-derivative ${\dot \phi}$ \cite{Bizon:2011gg} or other perturbations \cite{Dias:2011ss}. Rolling into an AdS region would have at least one non-zero ${\dot \phi}$ and a changing $\phi$, so we believe that the process of rolling into a classical AdS vacua is unstable. What happens next is unclear. The growth of $|{\dot \phi}| \rightarrow \infty$ in an AdS region indicating its instability means $\phi$ has to go somewhere else. It is likely that it has to roll out of the AdS region until it reaches either a Minkowski or a dS region. In the absence of a symmetry, a Minkowski vacuum is highly unlikely. (Following from the normalized probability distribution, we have
$$ \lim_{\epsilon \rightarrow 0} \int_0^{\epsilon}  P(\Lambda) = 0$$
even when $P(\Lambda)$ diverges at $\Lambda=0$.) This leaves us with any one of the many dS vacua in the low $\Lambda$ region of the landscape.

\subsection{Other Boson Mass Scales}

In the above string theory model, we have allowed each flux parameter to take a discrete set of values. A 2-form tensor field $C_2$ has a 3-form field strength $F_3=dC_2$ and its dual 
$F_7$ wrapping a 3-cycle yields a 4-form field strength $F_4$ in our 4-dimensional spacetime. It takes a discrete set of values, providing a constant contribution to the energy density,
$$V_i(F_4) \simeq  \frac{1}{2} F_{\mu \nu \rho \lambda}F^{\mu \nu \rho \lambda} = \frac{1}{2}(q_in_i)^2. $$
For example,
$b_i=q_in_i$, where $q_i$ depends on the embedding and the integer $n_i$ runs over the range of flux values. To be more precise, $n_i$ actually takes continuous values in an effective potential $V(n_i)$, where the minima of $V(n_i)$ sit at integer values of $n_i$. In the above analysis, we have assumed that the barriers between consecutive integer values are relatively high and so deviation from integer values are ignored. In actual cases, this means that the mass of $n_i$, namely $m_i^2 \simeq V''(n_i)$ are substantially bigger than those of the moduli considered above.  That is, besides the very light bosons, we do expect additional ones that are much heavier, though still much smaller than the string scale. Of course, the range of these masses depend on the details of the particular flux compactification.

\subsection{Cosmological Production}

Although the specific models discussed above may still be too simplistic for actual phenomenological studies, we can still comment on a few general issues related to cosmology. As pointed out in Sec. 1, recent investigations show that a very weakly coupled boson with mass $m \gtrsim 10^{-22}$ eV can be a good candidate for dark matter \cite{Hu:2000ke,Boehmer:2007um,Schive:2014dra}. A very low $\Lambda$ dS vacuum accompanied by light bosons may seem to fit the bill. However, when there are multiple light bosons, they may over-close the universe, especially if there are bosons with $m \gg 10^{-22}$ eV. When there are more than one light boson, the cosmological production can be quite involved.

The likely way to produce the bosons is via mis-alignment mechanism for axions \cite{Preskill:1982cy,Abbott:1982af,Dine:1982ah}. 
Let us review the scenario after inflation. The universe (or the inflaton) rolls down the landscape and moves towards the dS vacuum we are living in today. This rolling down follows a classical path, where damping takes place due to both the expansion of the universe and either decay and/or coupling to other fields. We expect it to follow close to the path of steepest descent. It may enter some local minima and, with enough kinetic energy, to roll out without being trapped.
At the last moment, it enters a local minimum and does not have enough energy to roll over the barrier it encounters; so it is trapped and will eventually settle in this local minimum. If it is moving along a particular axionic direction, it tends to oscillate along that direction around the minimum, producing non-relativistic axions via the misalignment mechanism. Fields along other moduli directions perpendicular to this direction will tend not to be produced, or little is produced. In general, rolling down the potential along a particular direction produces a linear combination of axions and/or light bosons. On the other hand, the initial condition can be tuned in our quasi-homogeneous universe such that overclosure of the universe did not happen \cite{Linde:1987bx}. 

Since we have little knowledge of the potential at finite temperature, especially around the low dS vacua, we have little to say about the impact of these light bosons on the dark matter scenario. Further study shall yield valuable constraints on the string theory scenario.

\section{Conclusion and Remarks}

The string theory models studied in this paper are admittedly relatively simple. Nonetheless, they incorporate known stringy properties in a consistent fashion so they are non-trivial enough for us to learn about the structure and dynamics of flux compactification in string theory. They clearly illustrate that a statistical preference for a very small physical $\Lambda$ in the cosmic landscape as a solution to the cosmological constant problem is a distinct possibility. This way to solve the cosmological constant problem bypasses the radiative instability problem. Associated with the very small $\Lambda$ are very light moduli masses. So this offers the possibility of having light bosons via statistical preference as well. It is important to point out that this solution or explanation is possible because of the existence of the landscape. Comparing to the earlier works \cite{Bousso:2000xa,Denef:2004ze,Denef:2004cf} where explicit interactions among the moduli and fluxes are not taken into account, we see that the statistical preference for a small $\Lambda$ (and at times some scalar masses) emerges only when couplings are included. Intuitively, in examining the models studied (albeit a rather limited sample),  more fluxes and moduli and more couplings among them tend to enhance or at least maintain the divergence of $P(\Lambda)$ at $\Lambda=0$. This is encouraging, since higher order corrections and more realistic (and so more complicated) models are very challenging to study. 

In terms of cosmology, one may wonder why the dark energy is so large, contributing to about $70\%$ of the content of our universe. However, from the fundamental physics point of view, the puzzle is why it is so small, when we know that the scale of gravity is dictated by the Planck scale $M_P$ which is so much bigger. Once we are willing to accept that the smallness of $\Lambda$ has a fundamental explanation like the statistical preference employed here, the question is again reversed. For example, in the viewpoint adopted here, we see that typical moduli mass scales are guided by $\Lambda$, not $M_P$. That is, some of the bosonic masses are expected to be very small. 

Once we accept that both $\Lambda$ and $M_P$ have their respective places in the theory (that is, generated by string theory dynamics ,with string scale $M_S$, not via fine-tuning), the presence of some intermediate mass scales such as the Higgs boson mass should not be so surprising. We see that the probability distribution $P(m^2)$ of bosonic mass $m^2$ does not peak at $m^2=0$ in the $\phi^3 / \phi^4$ model. In the string theory models, one envisions scenarios where some bosonic masses have a statistical preference for small values, but such preference is not as strong as that for $\Lambda$. So the Higgs mass $m_H =125$ GeV may fit in in such a scenario, thus evading the usual mass hierarchy problem for the Higgs boson. The scenario also offers the possibility that very light bosons can be present as the dark matter in our universe. In fact, any small number (e.g., the $\theta$ angle, light quark or neutrino masses in the standard electroweak model) in nature may be due to some level of a statistical preference without fine-tuning.

The string theory models considered in this paper are necessarily relatively simple, to allow semi-analytic studies. It will be important to consider more realistic versions (for example, the form of the K\"ahler potential and couplings among moduli) to see if such statistical preference for small $\Lambda$ and small bosonic masses are robust. 
In the search for the standard model within string theory, it may be fruitful to narrow the search of the three family standard model only in the region of the landscape where order of magnitude mass scales as well as $\Lambda$ come out in the correct range.

\section*{Acknowledgment}

We thank Thomas Bachlechner, Tom Broadhurst, Tiberiu Harko, Gary Horowitz, Lam Hui, Daniel Junghans, Gary Shiu, Plabo Soler, Yoske Sumitomo,  Andras Vasy, Ed Witten and Yi Wang for valuable discussions. This work is supported by the CRF Grant HKUST4/CRF/13G and the GRF 16305414 issued by the Research Grants Council (RGC) of  the Government of the Hong Kong SAR.

\appendix

\section{Characteristic equation} \label{AppA}

In finding the mass eigenvalues of the mass matrix (\ref{mreal}), the following matrix determinant identities are useful,
\begin{align}
\label{matrixa}
\det\begin{pmatrix} a & B^T \\ B & A \end{pmatrix} = (a-1)\det(A) + \det(A-BB^T), \\
\det \big(I_n + C_{n\times m}D_{m\times n} \big) =\det \big(I_m + D_{m\times n}C_{n\times m}\big).
\label{matrixf}
\end{align}

Suppressing the overall factor $4e^Kv^2$ in Eq.(\ref{mreal}) for the moment, the characteristic equation for the Hessian $H$ is simply the determinant $|H-\lambda I|$. Choosing $a$ in Eq.(\ref{matrixa}) to be $a=H_{11}-\lambda = 1+q -\lambda$, the determinant $\det(A)$ of the $n \times n$ matrix $A$ is given by 
 \begin{align}
\det (A) &=\det \big((4-\lambda)\delta_{ij} +n-4 +p_ip_j \big)= \det \big((4-\lambda)I_n + C_{n\times 2} D_{2 \times n} \big) \nonumber \\
&=(4-\lambda)^{n-2} \det \big(I_2 + D_{2 \times n}C_{n\times 2}\big), \nonumber \\
C^T&= \begin{pmatrix} p_1 & p_2 & ... & p_n \\ n-4 & n-4 & ... & n-4 \end{pmatrix}, \nonumber \\
D_{2 \times n} & =  \begin{pmatrix} p_1 & p_2 & ... & p_n \\ 1 & 1 & ... & 1 \end{pmatrix}.
\end{align}
This yields the first term in the characteristic equation (\ref{Charaeq}). Similar approach yields the second term in Eq.(\ref{Charaeq}) and the characteristic equation (\ref{Charaeqa}) for the axion masses. 

\vspace{4mm}

\bibliographystyle{utphys}
\bibliography{References}

\end{document}